\documentclass[printer]{aa}
\usepackage{graphicx,natbib}
\usepackage[latin1]{inputenc}
\usepackage[T1]{fontenc}
\bibpunct{(}{)}{;}{a}{}{,}
\usepackage{txfonts}
\include{def}
\usepackage{changebar}
\usepackage{bm}
\usepackage{units}
\usepackage{color}
\graphicspath{{./fig/}}
\usepackage{adjustbox}
\usepackage{longtable}

\begin{document}

\title{Metallicity effect and planet mass function in pebble-based planet formation models}

\author{N. Br\"ugger \inst{1}, Y. Alibert \inst{1}, S. Ataiee \inst{1,2}, W. Benz \inst{1}}
\offprints{N. Br\"ugger}
\institute{1.Physikalisches Institut, Universit\"at Bern, CH-3012 Bern, Switzerland \\
2. Institut f\"ur Astronomie \& Astrophysik, Universit\"at T\"ubingen, T\"ubingen, Germany\\
\email{natacha.bruegger@space.unibe.ch }
}

\abstract
{One of the main scenarios of planet formation is the core accretion model where a massive core forms first and then accretes a gaseous envelope. This core forms by accreting solids, either planetesimals, or pebbles.  A key constraint in this model is that the accretion of gas must proceed before the dissipation of the gas disc. Classical planetesimal accretion scenario predicts that the time needed to form a  giant planet's core is much longer than the time needed to dissipate the disc. This difficulty led to the development of another accretion scenario, in which cores grow by accretion of pebbles, which are much smaller and thus more easily accreted, leading to a more rapid formation.} 
{The aim of this paper is to compare our updated pebble-based planet formation model with observations, in particular the well studied metallicity effect.} 
{We adopt the Bitsch et al. 2015a disc model and the Bitsch et al. 2015b pebble model and use a population synthesis approach to compare the formed planets with observations.} 
{We find that keeping the same parameters as in Bitsch et al. 2015b leads to no planet growth due to a computation mistake in the pebble flux (Bitsch et al. 2017). Indeed a large fraction of the heavy elements should be put into pebbles ($Z_{\rm peb} / Z_{\rm tot} = 0.9$) in order to form massive planets using this approach. The resulting mass functions show a huge amount of giants and a lack of Neptune mass planets, which are abundant according to observations. To overcome this issue we include the computation of the internal structure for the planetary atmosphere to our model. This leads to the formation of Neptune mass planets but no observable giants. Reducing the opacity of the planetary envelope finally matches observations better.}
{We conclude that modeling the internal structure for the planetary atmosphere is necessary to reproduce observations.}

\titlerunning{New pebble-based planet formation}
\maketitle

\section{Introduction}
\label{introduction}

One of the main scenarios of planet formation is the core accretion model, which aims to explain, among other things, the formation of gas giants, believed to be the first object to form. 
These planets are massive and thus can have a huge impact on the formation of other planets in the system.
The idea of the core accretion model is that a massive  core forms first and then accretes a gaseous envelope. A key constraint in this model is that the accretion of gas must proceed before the dissipation of the gas disc. This implies that the core must become massive enough to accrete gas
in a few Myr (Haisch et al., 2001, Li \& Xiao, 2016).

In the classical accretion scenarios the core grows by accreting planetesimals, which are kilometer-sized objects and the time needed to form a  giant planet's core is much longer than the lifetime of the disc  (Pollack et al. 1996).
This difficulty led to the development of another accretion scenario, in which cores grow by accretion of pebbles, which are centimeter-sized bodies (Birnstiel et al. 2012). Pebbles themselves form by collisions of micro-meter sized 
dust embedded in the disc (Lambrechts \& Johansen 2014). Due to their small size, pebbles are more affected by gas drag and thus can be more easily accreted by the core (Lambrechts \& Johansen 2012),
resulting in a more rapid formation. 
While an appealing idea, this scenario needs to be confronted to observations to check wether it reproduces the trends seen in the data. 
We use the well studied metallicity effect which is fully established observationally for exoplanets and is reproduced by planetesimal-based formation models (Mordasini et al. 2009; Coleman \& Nelson, 2016a) to aim at testing this model, and present an updated pebble-based planet formation model which outcomes are comparable to observations.\\

This work is structured as follows: in section \ref{model} we set the theoretical basis that is used to perform our simulations. We present the disc model and its evolution, core growth by pebble and gas accretion, and the migration of the bodies. In section 
\ref{Formation_tracks} we compute formation tracks to disentangle the effects of growth and migration.
 In section \ref{population_synth} we present the impact of different initial conditions on the formation of planets, such as the starting time for the embryo and the metallicity of the disc, 
 using a population synthesis approach. 
 Considering a single planet per disc we then compare the results of our model with observations of exoplanets through mass functions. 
 In section \ref{Differentinternalstructure} we show the impact of having an internal structure for the planet's atmosphere and, adjusting some parameters, compare this updated model with observations.
 Finally, section \ref{metallicity}  discuss the metallicity effect and section \ref{conclusion} is devoted to discussion and conclusions.

\section{Theoretical model}
\label{model}

We present here the theoretical aspects behind our simulations. 
We define our disc model, its evolution and the core formation by pebble accretion. 
Migration is discussed and a comparison with previous work is provided as well.

\subsection{Protoplanetary disc model and evolution}
\label{discmodelevol}

We use  the disc model of Bitsch et al. (2015a) (hereafter B15a). In this model the disc structure depends on the accretion rate in the disc
 (assumed to be uniform) evolving with time as follows (Hartmann et al. 1998):
\begin{equation}
\log \left( \frac{\dot{M}}{M_{\odot} /yr} \right) = -8.00 - 1.40 \cdot \log \left( \frac{t + 10^5 yr}{10^6 yr} \right),
\label{discevolution}
\end{equation}
where $\dot{M}$ is related to the viscosity $\nu$ and the gas surface density $\Sigma_g$ via:
\begin{equation}
\dot{M} = 3 \pi \nu \Sigma_g = 3 \pi \alpha H^2 \Omega_k \Sigma_g,
\label{Mdotdisc}
\end{equation}
with $t$ being the time in years, $H$ the scale height of the disc, $\Omega_k$ the Keplerian frequency and $\alpha$ the viscosity parameter. The latter parameter 
influences the heating by viscous dissipation, and is set to $0.0054$ (B15a).\\

In order to compute the disc temperature, three different regimes are identified. In the inner disc (radial direction) the temperature is dominated by viscous heating, in the outer disc by stellar irradiation. Between the two regimes stands a cooler region that is heated by diffusion from the hotter inner part and the hotter outer part. This leads to three different power-law fits (see Appendix A of Bitsch et al. (2015a)).
Once the temperature is determined it can be linked to the different properties of the disc e.g $\Sigma$, P, H. 
For exemple the disc aspect ratio $\rm H/ \rm r$ is linked to the temperature via 
\begin{equation}
T = \left( \frac{H}{r} \right)^2 \frac{GM_{*}}{r} \frac{\mu}{R},
\label{eqTdisc}
\end{equation}
where $G$ is the gravitational constant, $M_{*}$ is the mass of the central star, $r$ the location in the disc, $R$ is the gas constant and $\mu$ is the mean molecular weight. \\
For a given  $\dot{M}$ and a specified temperature all quantities of the disc can be derived.

\subsection{Growth by pebble accretion}
\label{pebble_accretion}

During the evolution of the disc, embryos grow through the accretion of pebbles.  We use the pebble accretion model presented in Lambrechts \& Johansen (2012, 2014). In our simulations we assume that the embryos have already formed through the streaming instability, at a time that is a free parameter of the model.
 This means that we model the growth only above a certain mass which is given by the so-called pebble transition mass:
\begin{equation}
M_t = \sqrt{\frac{1}{3}} \frac{(\eta v_k)^3}{G \Omega_k},
\label{Mtransition}
\end{equation}
where $\eta = - \frac{1}{2} \left(\frac{H}{r} \right)^2 \frac{\partial ln P}{\partial ln r}$ and $v_k = \Omega_k r$, with $\Omega_k =  \sqrt{\frac{GM_*}{r^3}}$. 
This mass corresponds to the mass at which pebble accretion occurs within the Hill radius $r_H = r[M_c/(3M_*)]^{1/3}$ rather
than within the Bondi radius (Lambrechts \& Johansen 2012).
When $M_t$ is reached, the body starts to accrete efficiently with the 2D accretion rate (see Eq. \ref{dotM2D}).
As mentioned above, since we do not model the formation and growth until $M_t$, we assume a given starting time $t_{\rm ini}$ for the embryo. \\

In the outer regions of the disc a large reservoir of small dust slowly grows into larger inward-drifting pebbles (Lambrechts \& Johansen 2014). The location in the disc where the particles have just grown to pebble size is the pebble production line $r_g$, which is given by $t_{\rm growth}(r_g) = t_{\rm drift}(r_g)$ (outside the ice line). At this location the particles have reached the size for which their drift is more rapid than growth. As $r_g$ moves outwards with time, pebbles drift inward inducing a flux:
\begin{equation}
\dot M_{\rm pebbles} (r) = 2 \pi r_{\rm g} {d r_{\rm g} \over d t } \Sigma_{\rm dust} ( r_{\rm g} ) = 2 \pi r_{\rm g} {d r_{\rm g} \over d t } Z_{\rm peb} ( r_{\rm g} ) \Sigma_{\rm gas} ( r_{\rm g} ),
\label{eq_flux}
\end{equation}
where $\Sigma_{\rm dust}$ and $\Sigma_{\rm gas}$ are the surface densities of dust and gas respectively at $r_{\rm g}$, and $Z_{\rm peb}$ denotes the fraction of solids (with respect to the
amount of gas) that can turn into pebbles.  $Z_{\rm peb}$ is smaller than $Z_{\rm tot}$, the total fraction of solids (with respect to the amount of gas).
From this total fraction of solids $Z_{\rm tot}$, initially all in the form of dust, we assume that only a fraction forms pebbles (denoted $Z_{\rm peb}$) and the rest remains as dust (denoted $Z_{\rm dust}$). The latter influences the disc structure, the opacity of the disc and has an impact on migration via the surface density of the disc. On the other hand $Z_{\rm peb}$ regulates the amount of pebbles in the disc and thus the pebble surface density which impacts planet growth. $Z_{\rm tot}$ is assumed to be proportional to the metallicity of the star  $\rm[Fe/H]$ through:
\begin{equation}
\rm[Fe/H] = \log_{10} \left( \frac{Z_{\rm tot}}{Z_\odot} \right),
\end{equation} 
with $Z_\odot = 0.02$ (Anders \& Grevesse, 1989).\\

As the growth radius $r_{\rm g}$ moves outwards, the newly formed pebbles drift from the outer disc towards the star. At some point they cross the water ice line, which is the place in the disc where the temperature is equal to 170 K (see however Burn et al. in prep). Since pebbles are mainly composed of ices and silicates, when they reach the ice line, the ice sublimates releasing the trapped silicates causing their destruction (Morbidelli et al. 2015). 
In our model, a reduction of the flux of pebbles (50\%) is included at the water ice line.
However $\rm Z_{tot}$ and $\rm Z_{peb}$ are assumed to be independant of the semi-major axis throughout the disc evolution because the pebble production line moves quickly outwards far from the snow lines of volatile molecules such as $\rm H_2O$ , $\rm NH_3$ and $\rm CO_2$. 
Since we do not model the early stages of the disc evolution (assuming a starting time $\rm t_{ini}$ for the planets), the accreted pebbles are already beyond the snow lines when the embryo is inserted.
 By reducing the flux by 10\% at the $\rm CO_2$ line and 5\% at the $\rm NH_3$ line (based on Marboeuf et al. 2014b and Thiabaud et al. 2014), our results are unchanged.\\

The accretion rate of the pebbles onto the core is separated into two modes depending on the mass of the planet.
If the planet is small and its Hill radius is smaller than the scale height of pebbles $H_{\rm peb}$  then pebble accretion proceeds in the 3D mode (Bitsch et al. 2015b, hereafter B15b) :
\begin{equation}
\dot{M}_{c,3D} = \dot{M}_{c,2D} \left( \frac{\pi (\tau_f / 0.1)^{1/3} r_H}{2 \sqrt{2 \pi} H_{\rm peb}} \right).
\label{dotM3D}
\end{equation}
Then if $r_H$ becomes larger than  $H_{\rm peb}$, the accretion proceeds in 2D mode (Morbidelli et al. 2015): 
\begin{equation}
\dot{M}_{c,2D} = 2 \left( \frac{\tau_f}{0.1} \right)^{2/3} r_H v_H \Sigma_{\rm peb},
\label{dotM2D}
\end{equation}
where $\tau_f$ is the Stokes number depending on the size of the pebbles, $v_H= \Omega_k  \times r_H$ is the Hill speed and $\Sigma_{\rm peb}$ is the pebble surface density given by
\begin{equation}
\Sigma_{\rm peb} = \sqrt{\frac{2 \dot{M}_{\rm pebble} \Sigma_{\rm gas}}{\sqrt{3} \pi \epsilon_p a_p v_K}}.
\label{Sigmapeb}
\end{equation}
$\dot{M}_{\rm pebble}$ is given by Eq. \ref{eq_flux}, $\epsilon_p$ is the coagulation efficiency between pebbles and is assumed to be $0.5$ (Lambrechts \& Johansen 2014) and $a_p$ is the semi-major axis of the planet.\\
 The transition between 3D and 2D mode occurs when $H_{\rm peb} < \left( \frac{\pi (\tau_f / 0.1)^{1/3} r_H}{2 \sqrt{2 \pi}} \right)$ (B15b).\\
 
Outside the ice line, the pebble size is given by $t_{\rm growth}(r_g) = t_{\rm drift}(r_g)$, as explained above. However, inside the ice line, this assumption does not hold anymore. 
The size of the accreted pebbles inside the ice line is indeed much smaller ($\sim$ chondrule size (Morbidelli et al. 2015, Shibaike et al, in prep.)) than the one outside because of sublimation (Ida \& Guillot 2016).
Therefore, if a planet crossing the ice line has not reached the isolation mass, the accreted pebbles have a much lower Stokes number.
The surface density of pebbles should consequently be given by (Ormel \& Klahr 2010):
\begin{equation}
\Sigma_{\rm peb} = \frac{\dot{M}_{\rm peb}}{2 \pi r v_r},
\label{Sigmapeb_insideiceline}
\end{equation}
instead of Eq. (\ref{Sigmapeb}), which will influence the accretion rate. 
In Eq. \ref{Sigmapeb_insideiceline} the drift velocity $v_r$ is defined as (Weidenschilling 1977):
\begin{equation}
v_r = -2 \frac{\tau_f}{\tau_f^2 + 1} \eta v_K,
\end{equation}
where $\tau$ is given by the chondrule size.\\
Computing this reduction of the pebble size at the ice line we compare our accretion rate with formula 59 of Ida et al. 2016. 
They derive a general formula for the accretion rate of pebbles that is strongly influenced by the boundaries, such as the ice line or the transition from Epstein to Stokes drag regime.
Using Ida et al. 2016 formula we obtain a slightly smaller initial accretion rate, leading to a slower growth and therefore smaller planets.

Accreting pebbles, the core grows until it reaches what is called the isolation mass. This mass is defined as the mass required to perturb the gas pressure gradient in the disc, modifying the gas rotation velocity and thus halting the drift of pebbles (Ataiee et al. 2018, Bitsch et al. 2018):
\begin{equation}
M_{\rm iso} \propto 20 \left(\frac{H/R}{0.05}\right)^{3} \cdot M_{\oplus},
\label{Miso}
\end{equation}
where $M_{\oplus}$ is the mass of the Earth. When $M_{\rm iso}$ is reached the envelope of the planet contracts as long as $M_{\rm env} < M_{\rm core}$ and accretes gas following the accretion rate given by Piso \& Youdin 2014. Once $M_{\rm env} > M_{\rm core}$  a phase of rapid gas accretion starts (Lambrechts and Johansen 2014),
leading possibly to a gas giant. The accretion rate at this stage is given by Machida et al. 2010.

\subsection{Planet migration}
\label{planet_migration}

During the whole growth process forming planets migrate through the disc due to torques resulting from non-axisymmetric features in the gas induced by the presence of the planet. For low-mass planets that are still embedded in the disc, type I migration is described by Paardekooper et al. 2011:
\begin{equation}
\Gamma_{\rm tot} = \Gamma_L + \Gamma_c,
\end{equation}
where $\Gamma_L$ is the Lindblad torque and $\Gamma_c$ the corotation torque. In some regions where the corotation torque is bigger than the Lindblad one (i.e.  $\Gamma_{\rm tot}  > 0$), outward migration is possible. The sign and magnitude of the two torques arise from an asymmetry in the perturbed density distribution around the planet and wakes (Baruteau et al. 2014).\\

Planets that have reached their isolation mass start to accrete gas and grow rapidly until they finally open a gap in the disc. The criterion for gap opening is the one of Crida et al. 2006:
\begin{equation}
 \frac{3}{4} \frac{H}{r_H} + \frac{50}{q \Re} \leq 1,
\end{equation} 
where $q = M_{\rm planet} / M_{\rm star}$ is the mass ratio of the planet relative to the central star, $r_H$ is the Hill radius and $\Re$ is the Reynolds number defined by $\Re = \frac{r_p^2 \Omega_K}{\nu}$, with $r_p$ being the location of the planet in the disc and $\nu$ the viscosity.

If the criterion is fulfilled, the planet is massive enough to open a gap and migrates in type II regime. Type II migration is slower than type I, which potentially saves giant planets from ending up in the central star (Baruteau et al. 2014). We assume that the type II migration timescale of the planet is a function of the viscous timescale of the disc (Mordasini et al. 2009a) : $\tau_{II} = \frac{r_p^2}{\nu_{\rm visc}} \times \rm max(1, \frac{M_p}{4 \pi \Sigma_{\rm gas} r_p^2})$.

\subsection{Comparison with previous work}
\label{thebug}

In order to test our implementation of the equations presented in B15b, we attempted to reproduce their results.
First we compared the gas disc model with the results of Bitsch et al. 2015a, and found very good agreement\footnote{Note
that there is a typo in the formula of Bitsch et al. 2015a, and the formulas A17 of their annexe
must be corrected, the term in the second line multiplying the two terms of the first line, and not only the second one.}.

\begin{figure}
\hspace{0cm} \includegraphics[height=0.35\textheight,angle=0,width=0.35\textheight]{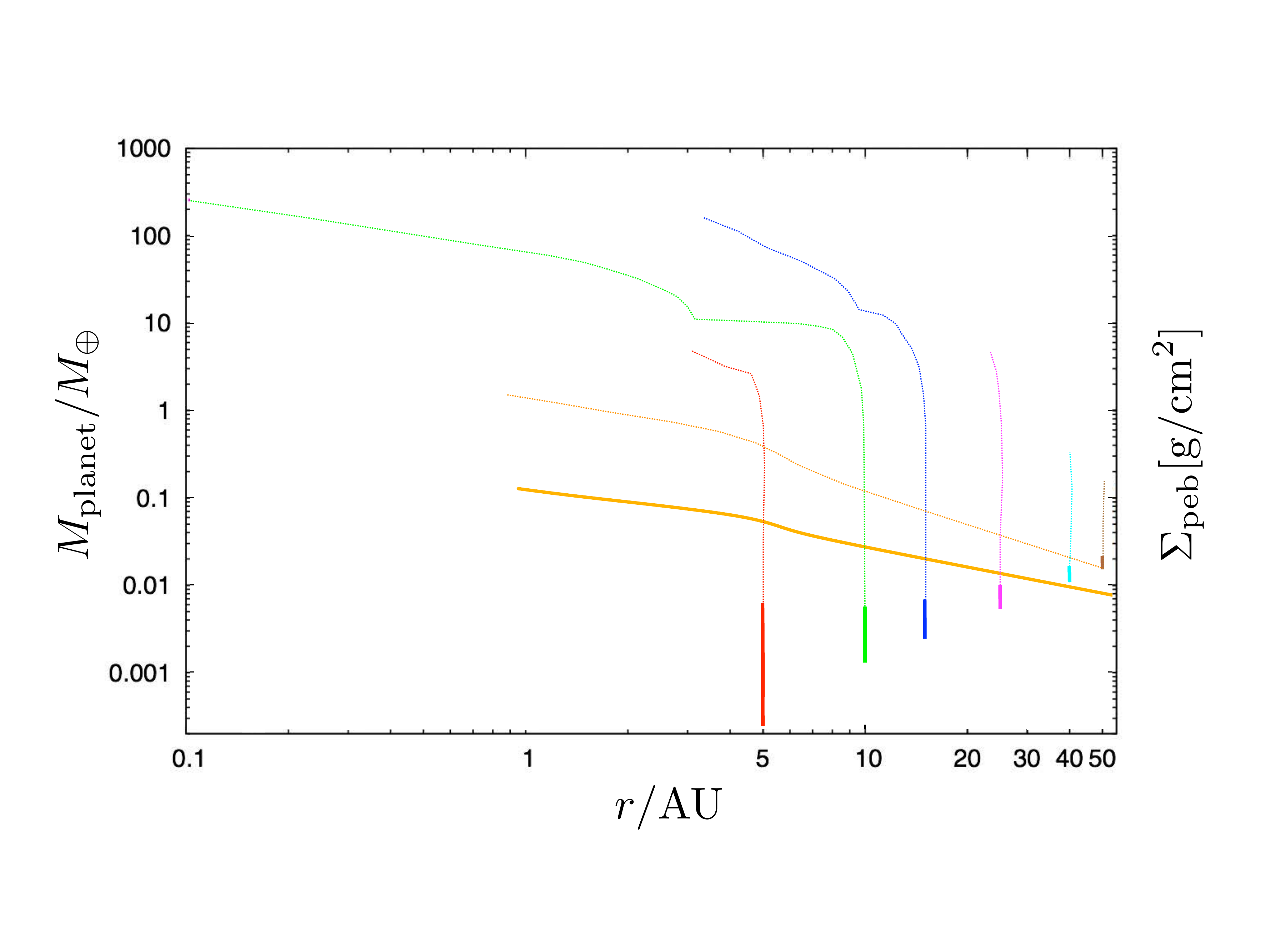}
  \caption{Comparison between the results of B15b (thin lines) and our results (thick lines). The lines starting vertically are the formation
  tracks of planets starting at 5, 10, 15, 25, 40, and 50 AU. The orange lines are the surface density of pebbles at 2 Myr in the two models.
  As in B15b, the planets are assumed to start their formation at 2 Myr.}
  \label{withoutbug}
\end{figure}

We then aimed at reproducing the growth of planets by pebble accretion. For this, we considered the same parameters
 as those quoted in B15b, and computed the formation of planets starting
at different locations in the disc, and at a time of 2 Myr (their Fig. 2). As explained in section \ref{pebble_accretion}, the amount of solids is split into the solids forming pebbles $Z_{\rm peb}$ and the ones remaining as dust $Z_{\rm dust}$. B15b uses $Z_{\rm peb} = 2/3 * Z_{\rm tot}$ and $Z_{\rm dust} = 1/3 * Z_{\rm tot}$.
The results of our model are shown in Fig \ref{withoutbug},
and it is immediately clear from this that our model (thick lines) \textit{does not} reproduce the results of B15b (thin lines). An interesting comparison can also be made using the surface density of pebbles at 2 Myr, also shown
in Fig. 2 of B15b. Comparing to that obtained in our case (see the yellow line in Fig. \ref{withoutbug}) and the one of B15b, we see
that the surface density of pebbles is much smaller in our simulation compared to the one of B15b.

\begin{figure}
\hspace{0cm} \includegraphics[height=0.35\textheight,angle=0,width=0.35\textheight ]{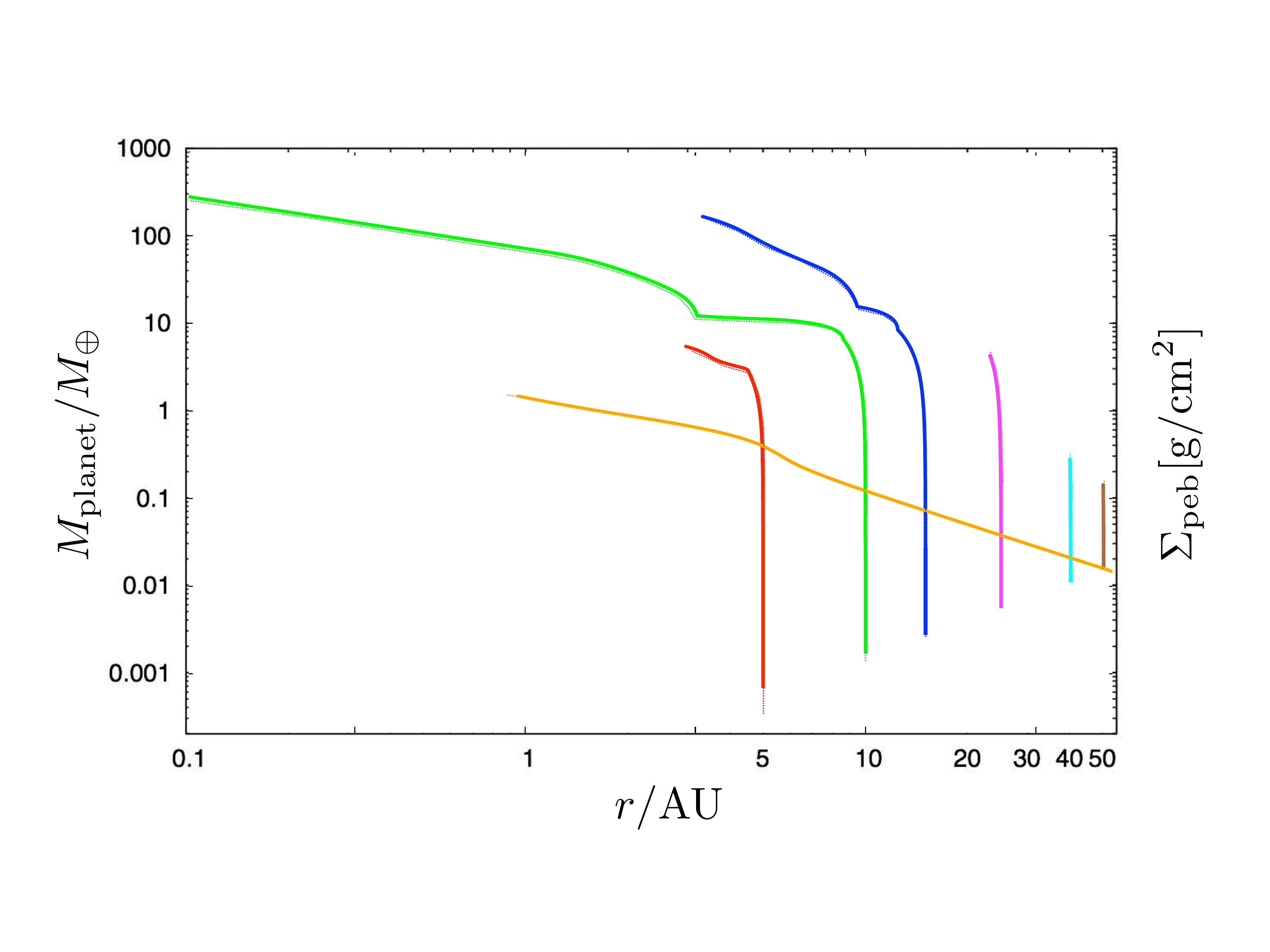}
  \caption{Same as Fig. \ref{withoutbug} but using Eq. \ref{eq_flux_bug} to compute the flux of pebbles. The match with the results of B15b is much better
  (the thin lines are barely visible but there is one such line below each of the thick lines).}
  \label{withbug}
\end{figure}

We then compute the same model, but now changing Eq. \ref{eq_flux} to the following equation:
\begin{equation}
\dot M_{\rm pebbles} (r) =  2 \pi r_{\rm g} {d r_{\rm g} \over d t } Z_{\rm peb} ( r_{\rm g} ) \Sigma_{\rm gas} ( r )
\label{eq_flux_bug}
\end{equation}
This equation is obviously incorrect because $\Sigma_{\rm gas}$ is taken at the local position in the disc $r$ and not at $r_g$. 
Surprisingly, in this case, the results were much closer to the ones of B15b.
We show this comparison in Fig. \ref{withbug}.

\begin{figure}
\hspace{0cm} \includegraphics[height=0.35\textheight,angle=0,width=0.35\textheight]{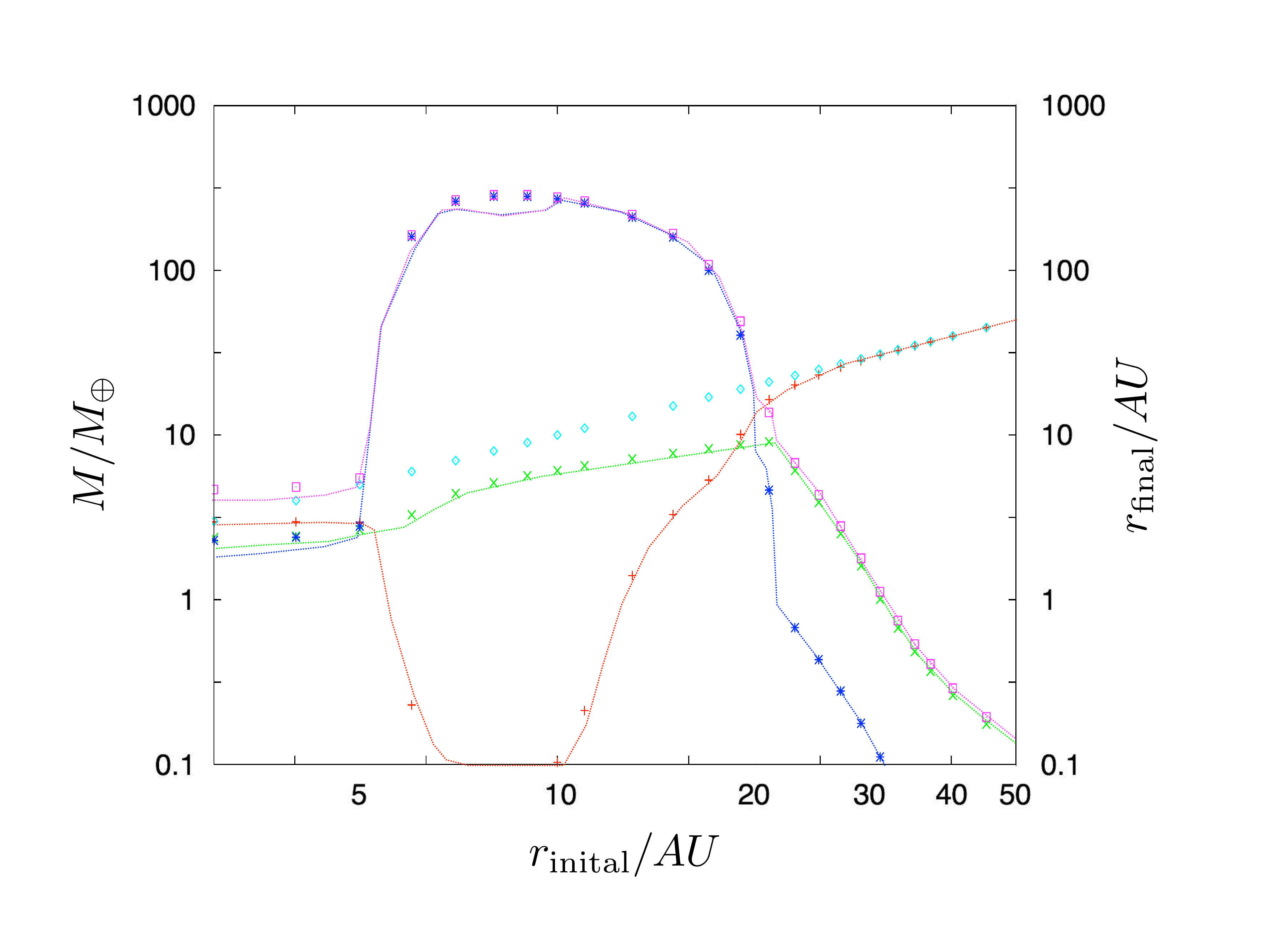}
  \caption{Final core mass (blue), final envelope mass (green), final total mass (magenta), and final location (red) of planets as a function of their
  initial location. The lines are the results of B15b, whereas the points with different symbols are our results. The cyan symbols are the initial location
  of planets. The pebble flux is given by Eq. \ref{eq_flux_bug}.}
  \label{fig3_withbug}
\end{figure}

In order to further check our results and compare them to B15b, we compute the core mass, envelope mass, total mass,
and final location of planets, as a function of their initial location, and compare our results with Fig. 3 of B15b. Using Eq. \ref{eq_flux_bug}  to compute the pebble flux, we obtain results very similar to those of B15b (see Fig. \ref{fig3_withbug}), whereas using the correct
 version of the equation (Eq. \ref{eq_flux}), the final planet mass is much smaller for all the starting locations. We then contacted B. Bitsch who confirmed to us that Eq. \ref{eq_flux_bug} instead of Eq. \ref{eq_flux} was indeed used in their code (private communication, Bitsch et al. 2017).

 As can be seen by comparing Figs. \ref{withoutbug} and \ref{withbug},
the effect on the growth of planets is dramatic: no planets grow when the pebble flux is computed with the correct equation, at least for the parameters
assumed in B15b.

\section{Formation tracks}
\label{Formation_tracks}

As explained in the previous section, using B15b parameters and the correct equation for the accretion rate (Eq. \ref{eq_flux}), do not result in planet formation. 
We thus have to adjust our model.
We changed the amount of solids used to form pebbles ($\rm Z_{\rm peb}$) and dust ($\rm Z_{\rm dust}$). 
This section shows the influences of both of them.
Several starting times for the embryos were tested (see section \ref{population_synth}) and since the conclusions were unchanged we only present in this chapter results for $\rm t_{\rm ini} = 1.5$ Myr.

\subsection{Changing $ Z_{\rm peb}$ and $ Z_{\rm dust}$ while keeping $  Z_{\rm tot}$ constant}
\label{ChangeZpebZdust_keepZtotconst}

\begin{figure}
\hspace{0cm} \includegraphics[height=0.35\textheight,angle=0,width=0.35\textheight]{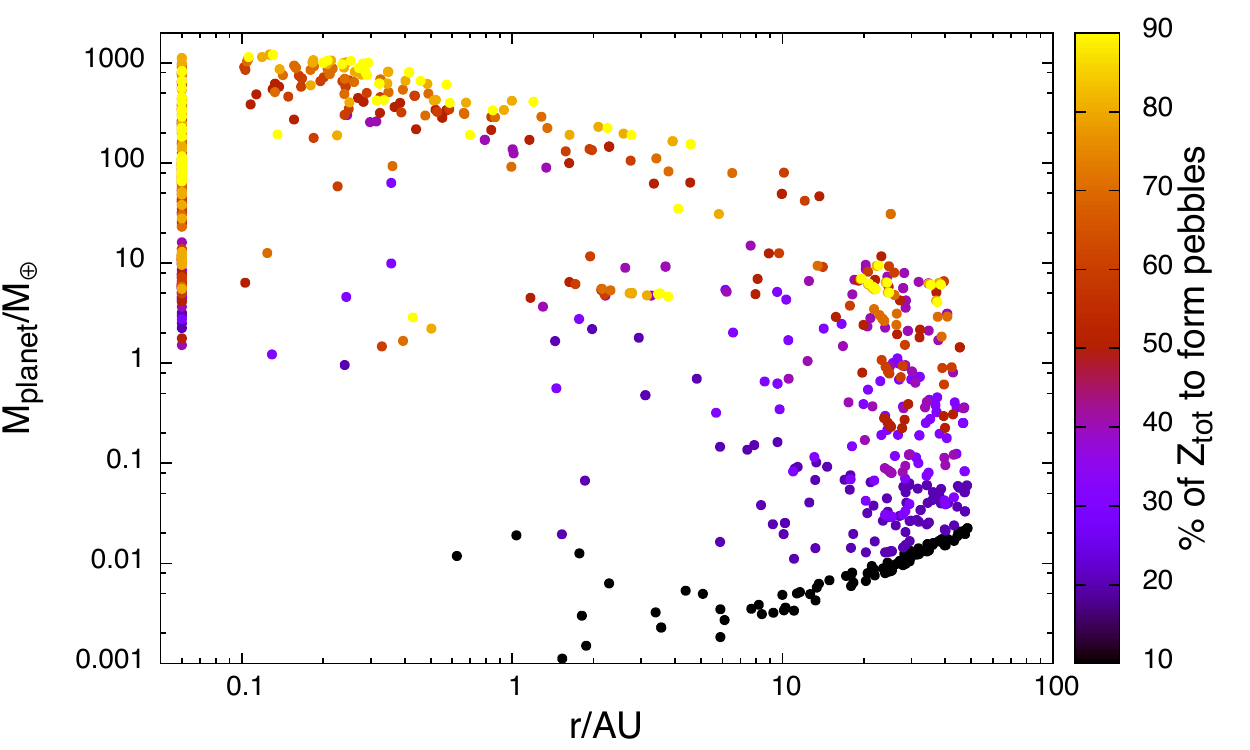}
  \caption{Mass of the planets as a function of their final location for a fixed value of $Z_{\rm tot} = 0.055$.  The starting time of the embryos is $\rm t= 1.5$ Myr with a linear distribution of their initial location (see discussion in section \ref{Initialconditions}). The color code gives the percentage of solids that is used to form pebbles ($Z_{\rm peb}$). The remaining percents of the solids is the dust in the disc ($Z_{\rm dust}$).}
  \label{Mvsa_for_fixedZtot}
\end{figure}

From the total amount of solids $\rm Z_{\rm tot}$, we vary the amount we consider to form pebbles and the one that remains as dust, as explained in section \ref{thebug}. In B15b this ratio was $2/3$ of $Z_{\rm tot}$ into pebbles and $1/3$ of $Z_{\rm tot}$ into dust. To test the influence of this ratio we take the constant value $\rm Z_{\rm tot} = 0.055$ and split this amount between dust and pebbles with different percentages\footnote{We choose $Z_{\rm tot} = 0.055$, which is almost three times the solar value of $Z_{\rm sun} = 0.02$ (Anders \& Grevesse, 1989), because it is an average value compared to the ones we test (see section \ref{population_synth}).}. Fig. \ref{Mvsa_for_fixedZtot} represents the mass of the formed planets as a function of their final location for this test. The colour code expresses the percentage of solids that have been considered to form pebbles. For example black dots represent the final properties of planets assuming that $10 \%$ of $\rm Z_{\rm tot}$ form pebbles and the other $90 \%$ remains as dust. In this case the formed planets are very small ($\sim 10^{-2}M_{\oplus}$) and no massive planet forms. More generally, Fig. \ref{Mvsa_for_fixedZtot} shows that the more material we use to form pebbles, the bigger the planets. Indeed most of the planets of $ \sim 1000$ $M_{\oplus}$ are represented by light orange or yellow points, corresponding to $\rm Z_{\rm peb} \sim 80 - 90 \%$ of $Z_{\rm tot}$.\\

\begin{figure}
\hspace{0cm} \includegraphics[height=0.35\textheight,angle=0,width=0.35\textheight]{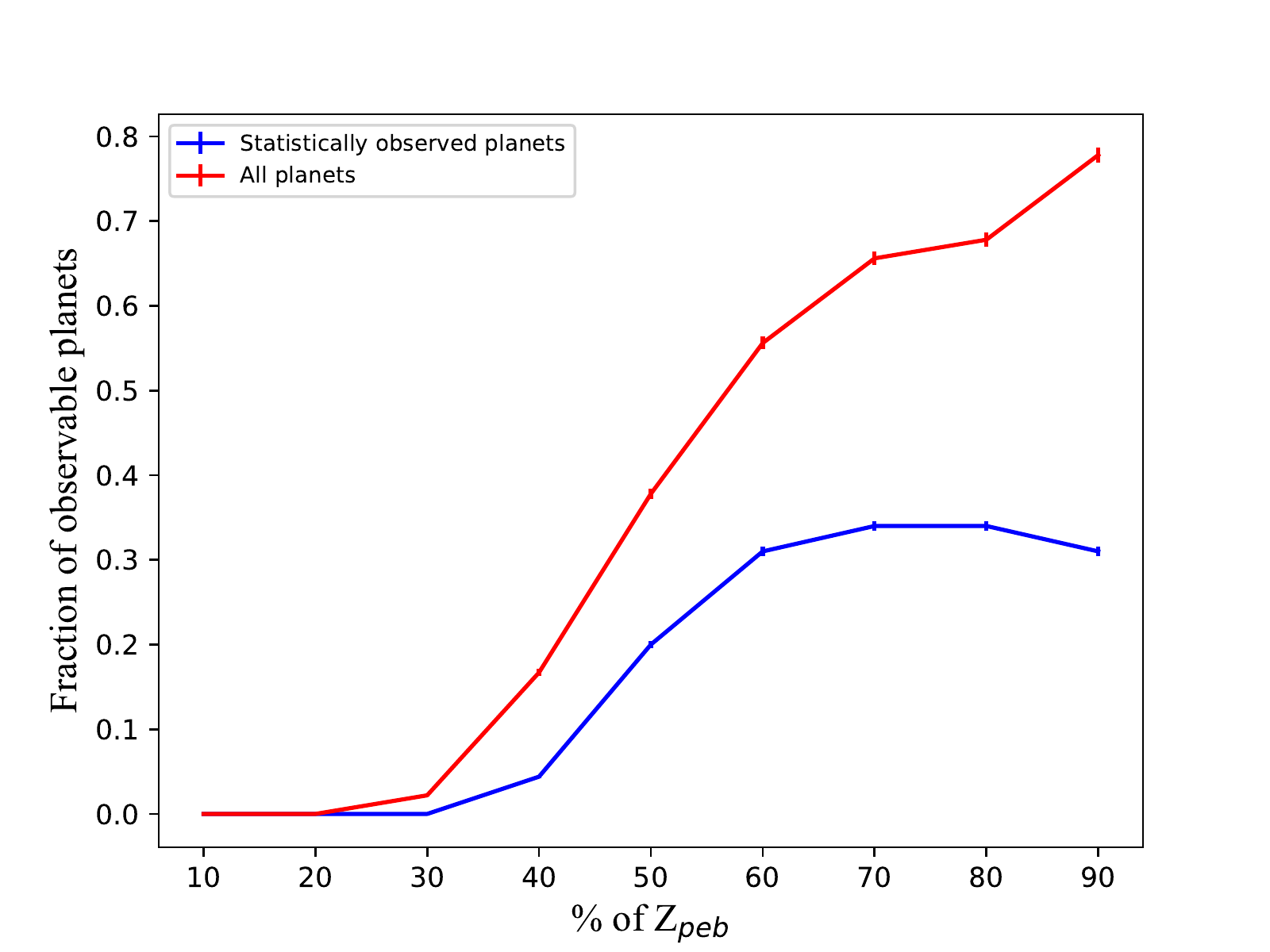}
  \caption{Fraction of planets with a radial velocity higher than 20 m/s as a function of the percentage of solids considered to form pebbles. The red curve gives the total amount of formed planets and the blue one gives the amount of formed planets beyond 0.1 AU. $Z_{\rm tot}$ is fixed and equal to 0.055.}
  \label{Zpeb_percentage_for_fixedZtot}
\end{figure}

Fig. \ref{Zpeb_percentage_for_fixedZtot} shows the percentage of planets with a radial velocity higher than 20 m/s depending on the amount of pebbles. 
Again here the total amount of solids is fixed: $Z_{\rm tot} = 0.055$. 
The percentage of planets is given for a radial velocity higher than 20 m/s to be consistent with Fig. \ref{KrvIS} and \ref{KrvB15} where we compare our results with those of Johnson et al. 2010. The red curve in Fig. \ref{Zpeb_percentage_for_fixedZtot} represents all the planets, including those piling up at $\rm r = 0.06$ AU which means that they are potentially lost into the star because they migrated to the inner edge of the disc (0.1 AU). The choice of piling the planet at $\rm r = 0.06$ AU, like in Fig. \ref{Mvsa_for_fixedZtot}, is arbitrary and avoid having a blank space in the figure between the planets that are still in the disc and the ones that fall into the star. 
Taking the red curve into account the most efficient percentage to form giant planets is the highest one shown here, $Z_{\rm peb} = 0.9 \times Z_{\rm tot}$.
On the other hand, the blue curve corresponds to all the planets beyond 0.1 AU, which is the inner edge of the disc. 
 In this case the percentage of giant planets slightly decreases for $Z_{\rm peb} = 0.9 \times Z_{\rm tot}$ compared to the $Z_{\rm peb} = 0.8 \times Z_{\rm tot}$ case. This is due to  too efficient migration if $Z_{\rm dust}$ is too small and thus planets fall into the star before accreting gas efficiently and possibly grow bigger. Therefore a smaller amount of $Z_{\rm dust}$, and thus a higher amount of $Z_{\rm peb}$, would not help the formation, nor the retention, of planets. Taking this into account and considering the two curves we conclude that the most efficient ratio to form giants is $Z_{\rm peb} = 0.9 \times Z_{\rm tot}$ even if some planets may migrate into the star.

\begin{figure}
\hspace{0cm} \includegraphics[height=0.35\textheight,angle=0,width=0.35\textheight]{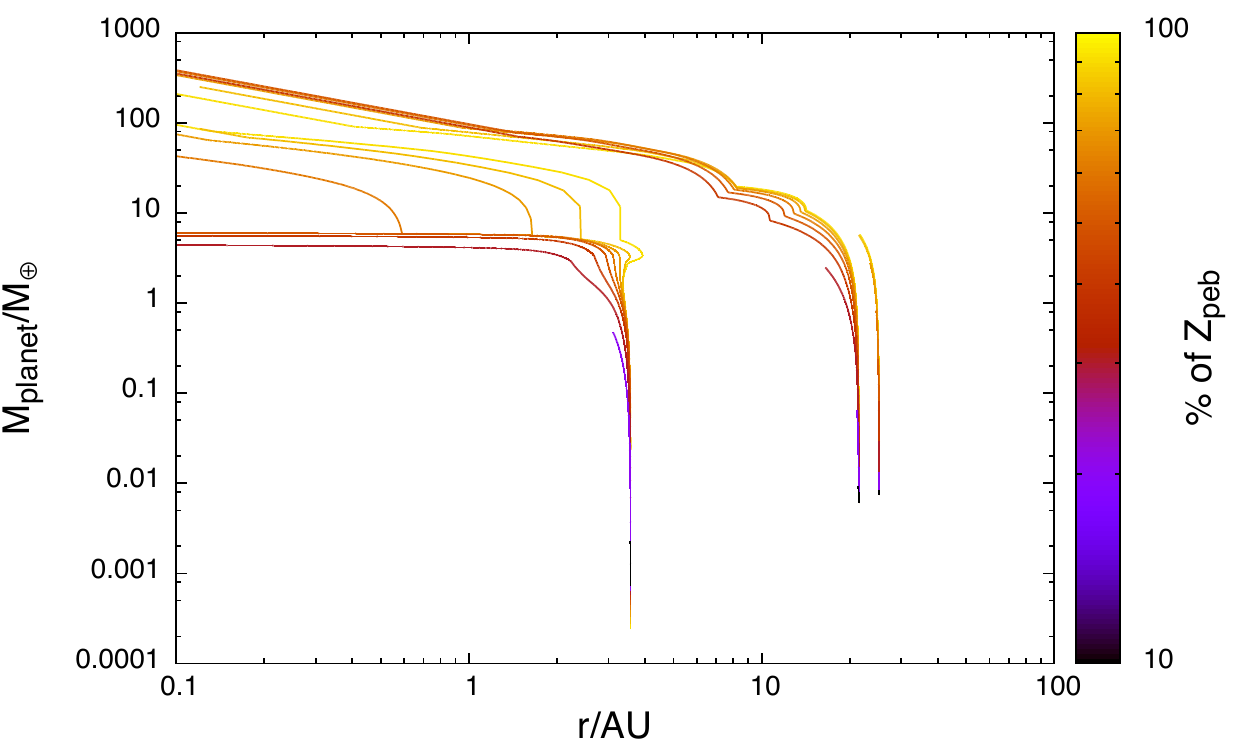}
  \caption{Growth tracks of planets for different starting locations. The colour code is the percentage of solids that is used to form pebbles. The remaining percents of the solids is the dust in the disc.}
  \label{Tracks_for_fixedZtot}
\end{figure}

Another argument in favour of this partition is observed in the growth tracks of the planets. Fig. \ref{Tracks_for_fixedZtot} represents growth tracks for planets starting at 3.5, 21.5 and 25 AU. Each starting location has been tested with a different ratio of pebbles and dust. We observe that for a too small ratio of pebbles and dust, regardless of the location, the planet does not manage to grow bigger than 1 $M_{\oplus}$. This is expected since there is not a lot of mass to accrete. Focusing on the planet starting at $\rm a_{\rm ini} = 3.5$ AU we see that there is a threshold value for which planets grow bigger than 10 $\rm M_{\oplus}$: an amount of 60 \% of pebbles seems necessary to form giant planets for close-in locations. Focusing on the other tracks, a threshold is not clearly distinguishable. However a large amount of pebbles favors the formation of giant planets. A fraction of 90 \% of pebbles ($\rm Z_{\rm peb}$) and 10 \% of dust ($\rm Z_{\rm dust}$) is therefore used in the rest of the paper (except in sections \ref{varyingZdustkeepingZpebcte} and \ref{varyingZpebkeepingZdustcte} which aim to show the influence of $\rm Z_{\rm peb}$ and $\rm Z_{\rm dust}$).

\subsection{Varying $ Z_{\rm dust}$ while keeping $ Z_{\rm peb}$ contant}
\label{varyingZdustkeepingZpebcte}

The amount of dust is then varied keeping the amount of pebbles constant to test the impact of the disc structure on the formed planets. 
Indeed $\rm Z_{ \rm dust}$ influences the disc structure which will have consequences not only for the migration of planets but also for the isolation mass (Eq. \ref{Miso}): the disc aspect ratio H (which appears in the $\rm M_{\rm iso}$ formula) is regulated by the temperature (as can be seen in Eq. \ref{eqTdisc}) which is influenced by the amount of dust $\rm Z_{\rm dust}$ (B15b).\\
The variation of $\rm Z_{\rm dust}$ modifies the total amount of solids ($\rm Z_{\rm tot}$) for each test. 
Table \ref{Tableau_Zpebconst_Zdustvary} labels the different cases we use. 
In order to compare the tracks for the different $\rm Z_{\rm dust}$ we choose a random initial location in the disc (here $a_{\rm ini} = 6.6$ AU) and look at the growth. \\


\begin{center}
\begin{table}
\centering
\caption{Amount of pebbles, dust and solids in the test for a constant $\rm Z_{\rm peb}$.}
\label{Tableau_Zpebconst_Zdustvary}
\begin{tabular}{|c|c|c|c|c|}
\hline
\textbf{$\rm Z_{\rm peb}$} & \textbf{$\rm Z_{\rm dust}$} & \textbf{$\rm Z_{\rm tot}$}  &\textbf{$a_{\rm final} / AU$} &  \textbf{$M_p / M_{\oplus}$} \\
\hline
0.05 & 0.001 & 0.051 & 2.99 & 1.76 \\
\hline
0.05 & 0.003 & 0.053 & 3.02 & 4.46 \\
\hline
0.05 & 0.005 & 0.055 & 3.68 & 4.89 \\
\hline
0.05 & 0.008 & 0.058 & 3.20 & 4.98 \\
\hline
0.05 & 0.01 & 0.06 & 2.70 & 5.49 \\
\hline
0.05 & 0.02 & 0.07 & 0.05 & 10.09 \\
\hline
0.05 & 0.03 & 0.08 & 0.05 & 55.98 \\
\hline
0.05 & 0.04 & 0.09 & 0.22 & 226.77 \\
\hline
0.05 & 0.05 & 0.1 & 0.26 & 231.60 \\
\hline
0.05 & 0.07 & 0.12 & 0.27 & 233.67 \\
\hline
\end{tabular}
\end{table}
\end{center}

\begin{figure}
\hspace{0cm} \includegraphics[height=0.35\textheight,angle=0,width=0.35\textheight]{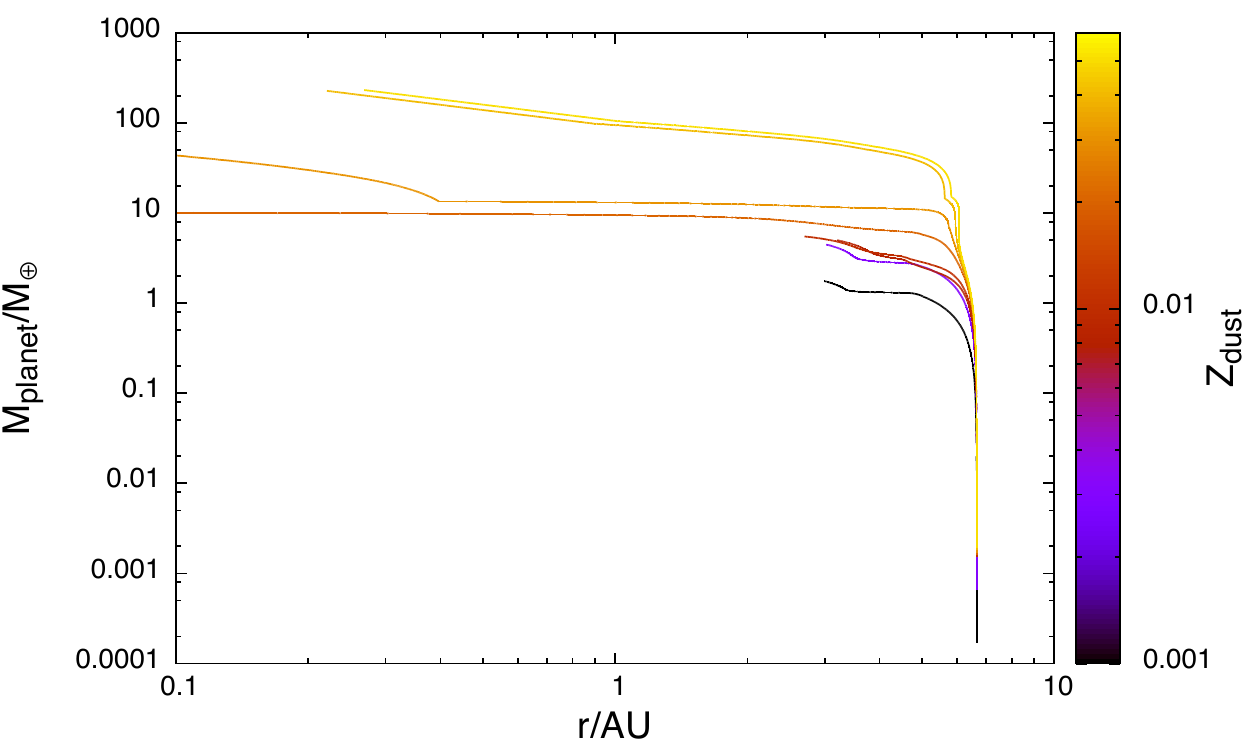}
  \caption{Growth tracks for planets starting at the same location for different $\rm Z_{\rm dust}$ and fixed $\rm Z_{\rm peb} = 0.05$. The colour code expresses the different $\rm Z_{\rm dust}$.}
  \label{Zdust_change_for_fixedZpeb}
\end{figure}

 Analysing Fig. \ref{Zdust_change_for_fixedZpeb} allows us to distinguish three different regimes.
The first one concerns the planets with $\rm Z_{\rm dust}$ between 0.001 and 0.01: they all reach their isolation mass but due to their small $\rm Z_{\rm dust}$ their $\rm M_{\rm iso}$ is too small to accrete gas efficiently (their $\rm M_{\rm iso}$ stands between $1.2$ $\rm M_{\oplus}$. and $3.1$ $\rm M_{\oplus}$). 
The planet with $\rm Z_{\rm dust} = 0.02$ also reaches its isolation mass but is big enough to accrete gas efficiently ($\rm M_{\rm iso} \sim 5.75$ $\rm M_{\oplus}$). However it does not succeed in opening a gap in the disc and thus undergoes type I migration until it falls into the star. The third regime contains the planets with $\rm Z_{\rm dust}$ between 0.03 and 0.05. They all reach their isolation mass and are big enough to accrete gas. The planet with $\rm Z_{\rm dust} = 0.03$ first migrates with type I migration and then becomes massive enough to open a gap in the disc and migrate with type II. The two other ones reach $\rm M_{\rm iso} \sim 8$ $\rm M_{\oplus}$ and enter in the type II migration regime rapidly which prevents them from falling into the star. \\
These tests emphasise the fact that the effect of the dust fraction is important for not only the disc structure, but also the growth of the planet, mainly via the impact on the isolation mass.

\subsection{Varying $ Z_{\rm peb}$ while keeping $ Z_{\rm dust}$ constant}
\label{varyingZpebkeepingZdustcte}

In order to test the influence of the amount of pebbles, keeping the disc structure constant, we run similar simulations, but now keeping a constant value of $Z_{\rm dust} = 0.005$, and varying $Z_{\rm peb}$  from 0.001 to 0.1 (see Table \ref{Tableau_Zdustconst_Zpebvary}). 
The major impact of the amount of pebbles $Z_{\rm peb}$ is on the surface density of pebbles and thus on the planet growth, as can be observed in  Table \ref{Tableau_Zdustconst_Zpebvary}, which summarises the parameters used for each of these simulations.\\
We conclude that increasing the amount of pebbles, the larger the planets can grow. However, as already mentionned, the amount of dust should not be too small otherwise the migration of the planets becomes too efficient.


\begin{table}
\centering
\caption{\label{Tableau_Zdustconst_Zpebvary} Amount of pebbles, dust and solids in the test for a constant $\rm Z_{\rm dust}$.}
\begin{tabular}{|c|c|c|c|c|c|c|c|c|} 
\hline
\textbf{$\rm Z_{\rm peb}$} & \textbf{$\rm Z_{\rm dust}$} & \textbf{$\rm Z_{\rm tot}$}  &\textbf{$a_{\rm final} / AU$} &  \textbf{$M_p / M_{\oplus}$} \\
\hline
0.001 & 0.005 & 0.006 & 6.65 & 0.00106 \\
\hline
0.003 & 0.005 & 0.008 & 6.65 & 0.00124 \\
\hline
0.007 & 0.005 & 0.012 & 6.65 & 0.00286 \\
\hline
0.01 & 0.005 & 0.015 & 6.64 & 0.00864 \\
\hline
0.03 & 0.005 & 0.035 & 3.61 & 3.91 \\
\hline
0.05 & 0.005 & 0.055 & 3.68 & 4.89 \\
\hline
0.08 & 0.005 & 0.085 & 3.51 & 5.74 \\
\hline
0.1 & 0.005 & 0.105 & 3.43 & 6.24 \\
\hline
\end{tabular}
\end{table}

\section{Population synthesis}
\label{population_synth}

Based on the previous section's discussion we set our nominal model to be the B15b approach (with the correct equation for the accretion rate) and the pebble formation efficiency at 90 \% with the remaining 10 \% being dust.\\
In order to compare the results of our model with observations, we need to take into account the wide existing spectrum of disc properties. We compute the migration of the planets following the formulae of Paardekooper et al. 2011 instead of those given in B15b. These two migrations are quite similar except for the fact that B15b uses a smothing function for the transition between the type I and type II migration regimes.

\subsection{Initial conditions}
\label{Initialconditions}

The initial conditions of our model can be split into two types of parameters: the ones set for a whole population and the ones varied in each simulation.\\
The former category includes the starting times of the planetary seeds and the total amount of solids.
The seeds, each of them having a mass given by Eq. \ref{Mtransition}, are inserted at the beginning of the evolution (0 Myr), after 0.5, 1, 1.5 and 2 Myr. 
As can be seen in Table \ref{Tableau_pourcent}, if the seeds are introduced too early in the simulation most of the planets end up in the star. Therefore we only focus on starting times of 1.5 Myr and 2 Myr.
For those two starting times, the total amount of solids is varied between 0.011 and 0.11.\\

\begin{figure}
\hspace{0cm} \includegraphics[height=0.35\textheight,angle=0,width=0.35\textheight]{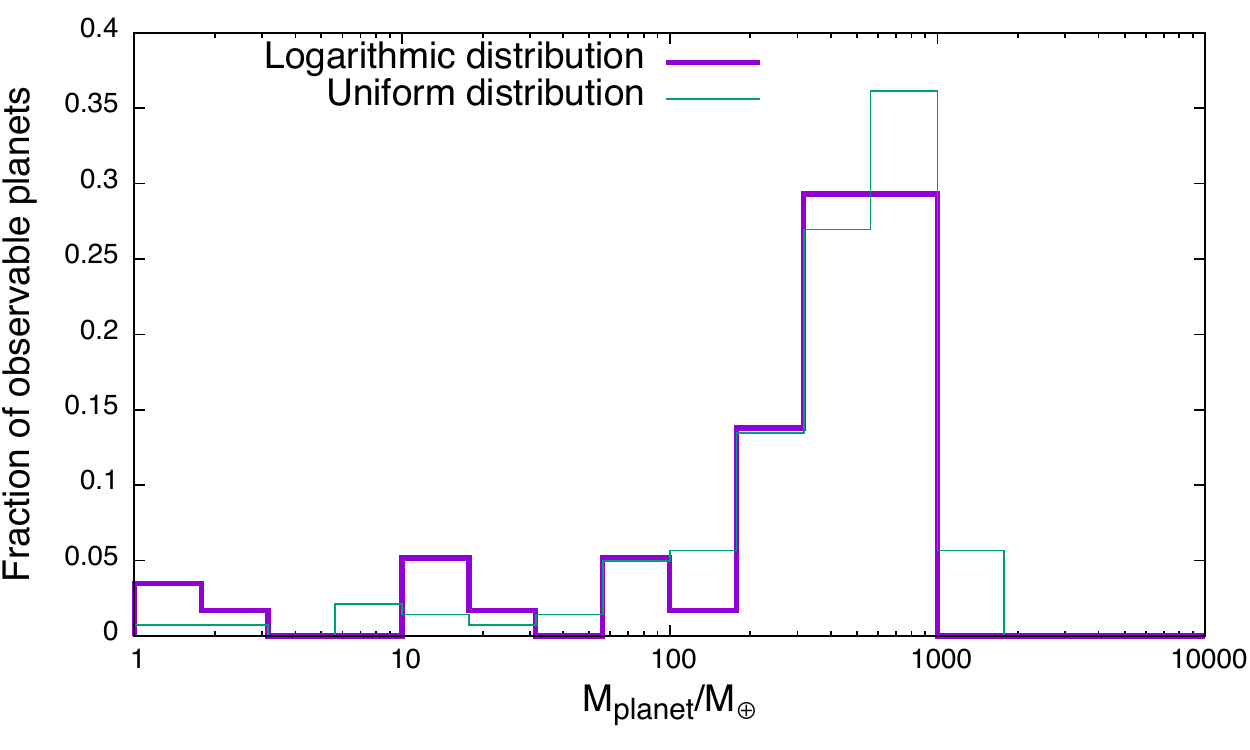}
  \caption{Mass function for planets starting with a uniform distribution of the initial locations compared to a logarithmic distribution.}
  \label{Massfunction_LOG_UNIFORM}
\end{figure}

Each simulation is modeling the growth of a single planet per disc with a random lifetime of this disc and initial location of the seed. 
The lifetime is important because it impacts the duration of the planet growth. 
In our model, it is randomly chosen between 2 and 5 Myr with a uniform distribution (Haisch et al. 2001b). 
The disc evolution stops when the lifetime is reached or when the planet has migrated into the star. 
Note that the evolution of the disc remains the same for the different liftetimes, only the final time changes.
The initial location distribution is assumed to be linear between 0.1 and 50 AU. 
As can be seen in Fig. \ref{Massfunction_LOG_UNIFORM} the function of mass of the formed planets using a linear or a logarithmic distribution look very similar. 
However using a logarithmic distribution the model will form more planets in the inner disc where the isolation masses are smaller, resulting in planets that will not reach the runaway gas accretion phase and may fall quickly into the star (Bitsch \& Johansen 2017).
We thus decide to use a linear distribution between 0.1 and 50 AU.


\begin{table}
\centering
\caption{\label{Tableau_pourcent} Percentages of planets ending in the star (closer than 0.1 AU) according to the different starting times of the seed.}
\begin{tabular}{|c|c|} 
\hline
Starting times & \% of planets ending in the star \\
\hline
0 Myr & 91 \% \\
\hline
0.5 Myr & 86 \% \\
\hline
1 Myr & 55 \% \\
\hline
1.5 Myr & 40 \% \\
\hline
2 Myr & 29 \% \\
\hline
\end{tabular}
\end{table}

\subsection{Mass- semi-major axis diagram}

\begin{figure}
\hspace{0cm} \includegraphics[height=0.35\textheight,angle=0,width=0.35\textheight]{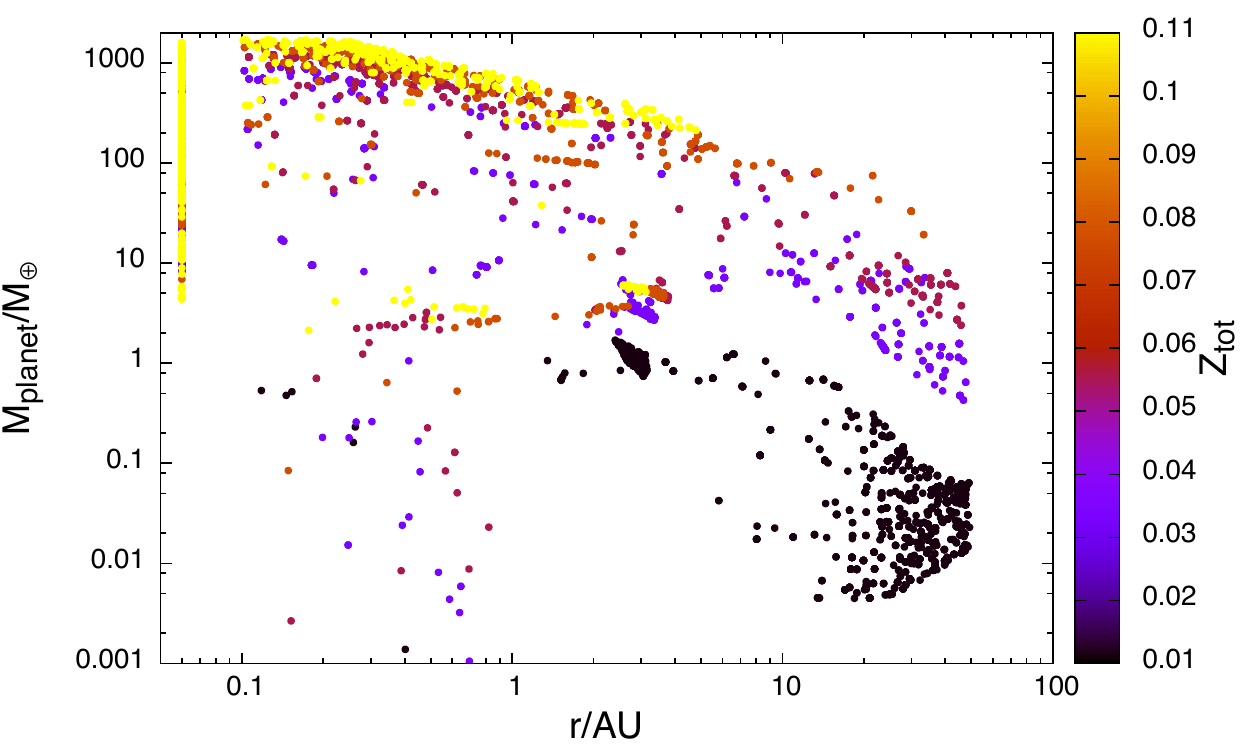}
  \caption{Mass of planets as a function of their final semi-major axis for seeds starting after 1.5 Myr of disc evolution. The color code represents the different amount of solids $Z_{\rm tot}$ (from which 90 \% will form pebbles and 10 \% will remain as dust.). The spacing between the black dots and the purple ones comes from the fact that we use discrete values of $\rm Z_{\rm tot}$.}
  \label{Mvsa_t1.5}
\end{figure}

\begin{figure}
\hspace{0cm} \includegraphics[height=0.35\textheight,angle=0,width=0.35\textheight]{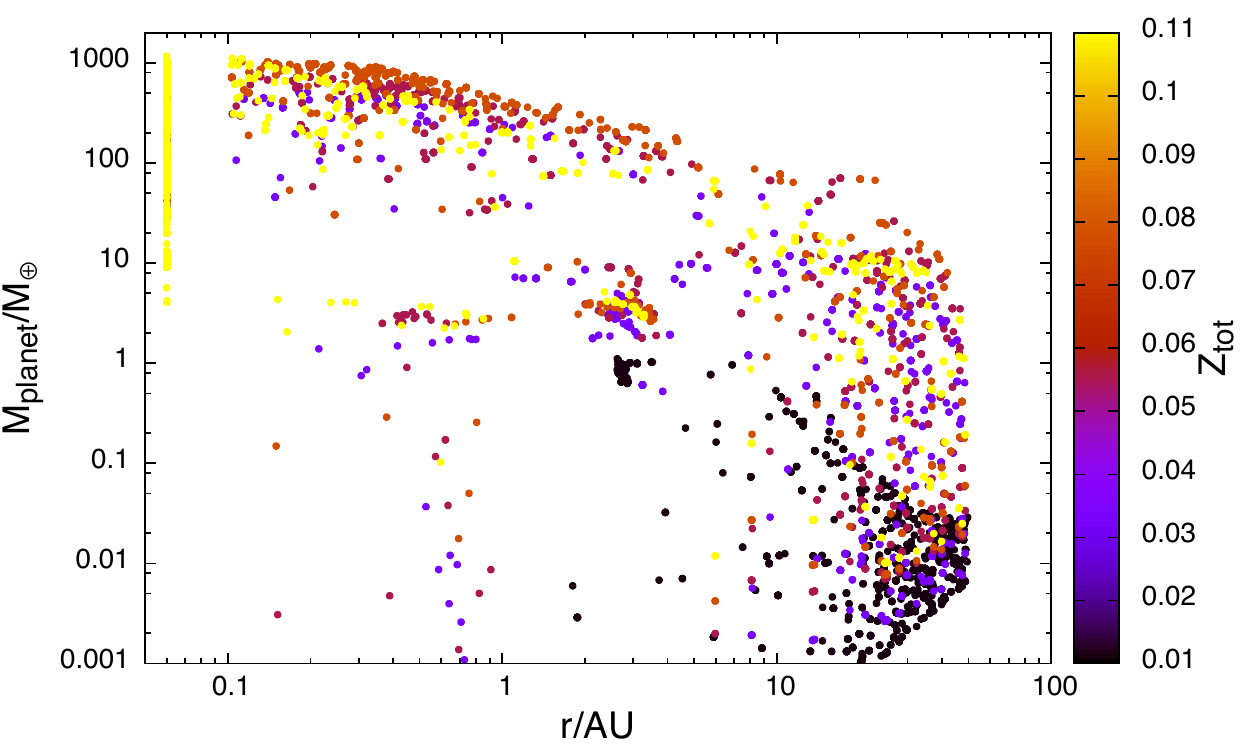}
  \caption{Mass of planets as a function of their final semi-major axis for seeds starting after 2 Myr of disc evolution (same as Fig. \ref{Mvsa_t1.5} but for $\rm t_{\rm ini} = 2$ Myr). The color code represents the different amount of solids $Z_{\rm tot}$.}
  \label{Mvsa_t2}
\end{figure}

Fig. \ref{Mvsa_t1.5} and \ref{Mvsa_t2} show the mass of the formed planets as a function of their final semi-major axis for different metallicities. 
The pile up at $\rm r = 0.06$ AU that can be observed in both figures represents the planets that might be lost into the star because they migrated to the inner edge of the disc (0.1 AU). 
Focusing on the final locations and masses of the planets we observe a region with a lack of planets between 1 and 3 AU for masses around 0.01-0.1 $M_{\oplus}$. This is an effect caused by the ice line.
The planets that reach 100 to 1000 $M_{\oplus}$ appearing between 0.1 and 1 AU are the ones that come from further out in the disc and that have migrated inwards.
The smaller planets within this region are the ones with initial locations below 1 AU.
They remain near their initial location since planets start migrating once they reach the isolation mass.
We see some Earth-mass planets forming interior to 1 AU but since we only have one planet per system, we cannot really rely upon them (Alibert et al. 2013).
We leave interactions between planets for future work.
Note that the planet distribution near the inner edge of the disc could be largely changed due to new type II migration models (Kanagawa et al. 2018, Robert et al. 2018 and Ida et al. 2018). These models show that for low viscosity cases the migration towards the star would be slowed down. This is however not considered in the present paper but will be treated in future work.\\

In Fig. \ref{Mvsa_t1.5}, the seeds are placed in the disc after 1.5 Myr of evolution. 
A clear tendency emerges: an increased pebble density favors the formation of more massive planets.
For a small amount of solids such as $Z_{\rm tot} = 0.011$ (meaning that $Z_{\rm peb} = 0.01$ and $Z_{\rm dust} = 0.001$) planets do not grow larger than 1 $\rm M_{\oplus}$. 
Most of them even remain below $0.1$ $\rm M_{\oplus}$ and beyond 10 AU due to a lack of material for those embryos to accrete and grow bigger. 
However, if they nevertheless manage to reach a certain mass, the amount of dust is very small and their migration is too efficient to prevent them from falling into the star (as discussed in section \ref{Formation_tracks}). 
These results are valid for $Z_{\rm tot} =0.033$ as well. 
However, for higher amounts, like $Z_{\rm tot} = 0.055$, $Z_{\rm tot} = 0.077$ and $Z_{\rm tot} = 0.11$, 100 to 1000 $M_{\oplus}$ planets appear, mostly between 0.2 AU and 5 AU. \\

For seeds starting after 2 Myr, the distribution of planets is not too dissimilar to the one with $ \rm t_{ \rm ini} = 1.5$ Myr. 
We still observe that with more solids in the disc (i.e. higher $Z_{\rm tot}$), planets are more massive. 
However the distinction between the different amounts of solids is less obvious than in Fig. \ref{Mvsa_t1.5}.

\subsection{Comparison of mass functions}
\label{Massfunction}

\begin{figure}
\hspace{0cm} \includegraphics[height=0.35\textheight,angle=0,width=0.35\textheight]{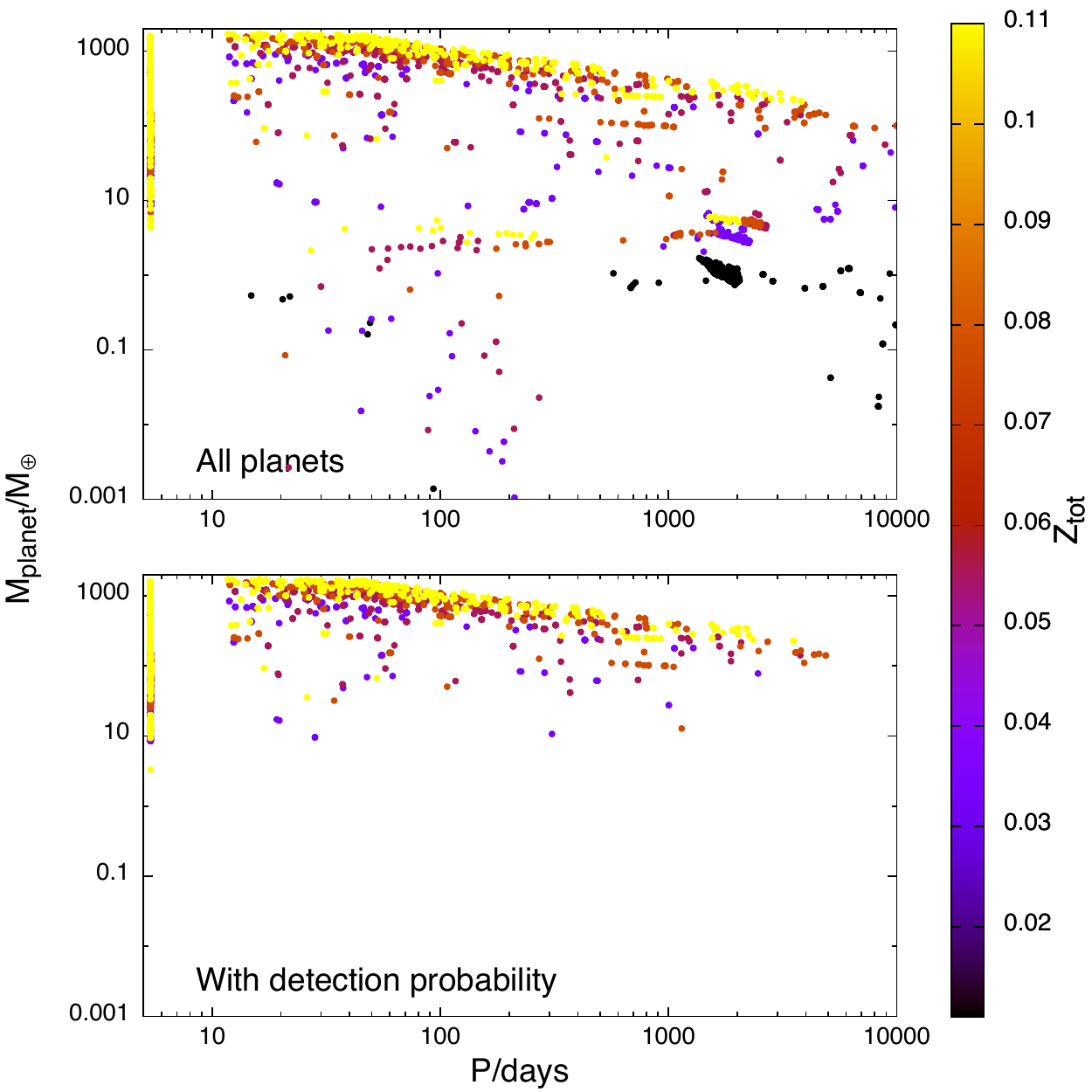}
  \caption{Mass of the planets as a function of their period for starting time of the embryos of $\rm t = 1.5$ Myr. The plot on the top represents all planets formed and the bottom one the planets that are considered to be observed once the detection probability of Mayor et al. 2011 is applied. The color code expresses the total amount of solids.}
  \label{multiplot_MvsP_1.5}
\end{figure}

\begin{figure}
\hspace{0cm} \includegraphics[height=0.35\textheight,angle=0,width=0.35\textheight]{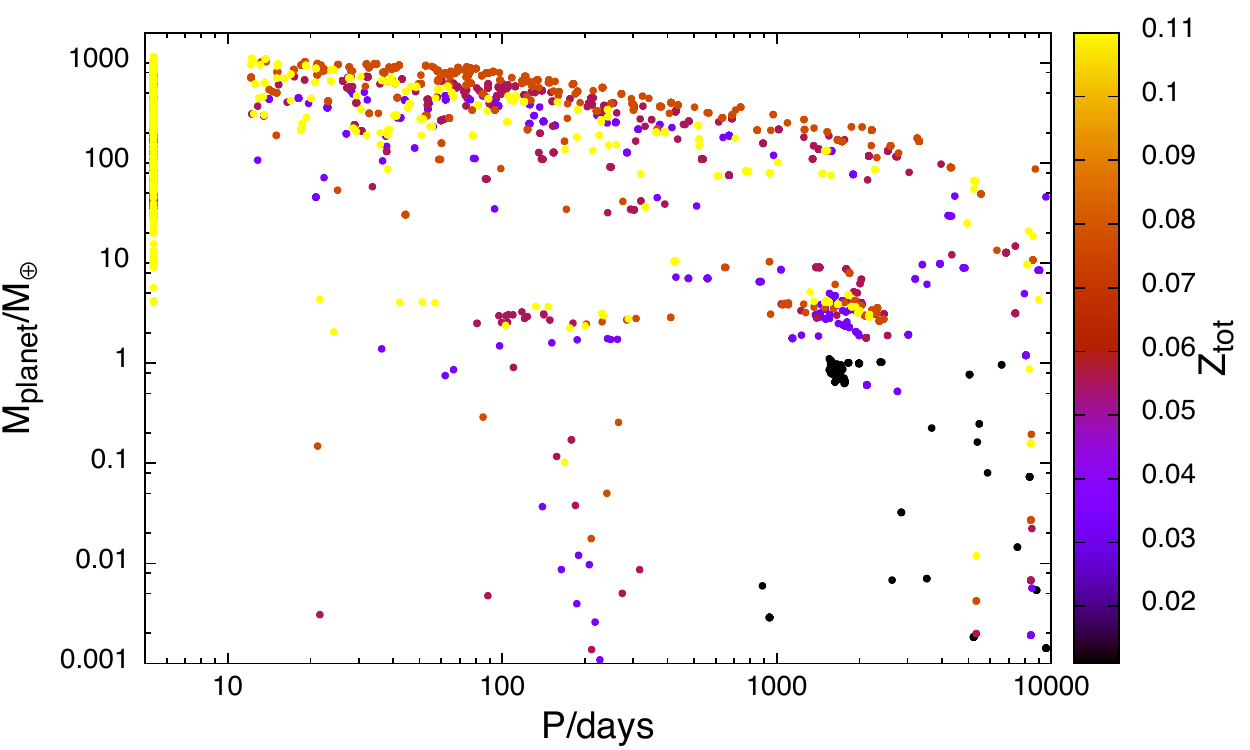}
  \caption{Mass of all the formed planets as a function of their period for starting time of the embryos of $\rm t = 2 $ Myr.}
  \label{multiplot_MvsP_2}
\end{figure}

\begin{figure}
\hspace{0cm} \includegraphics[height=0.35\textheight,angle=0,width=0.35\textheight]{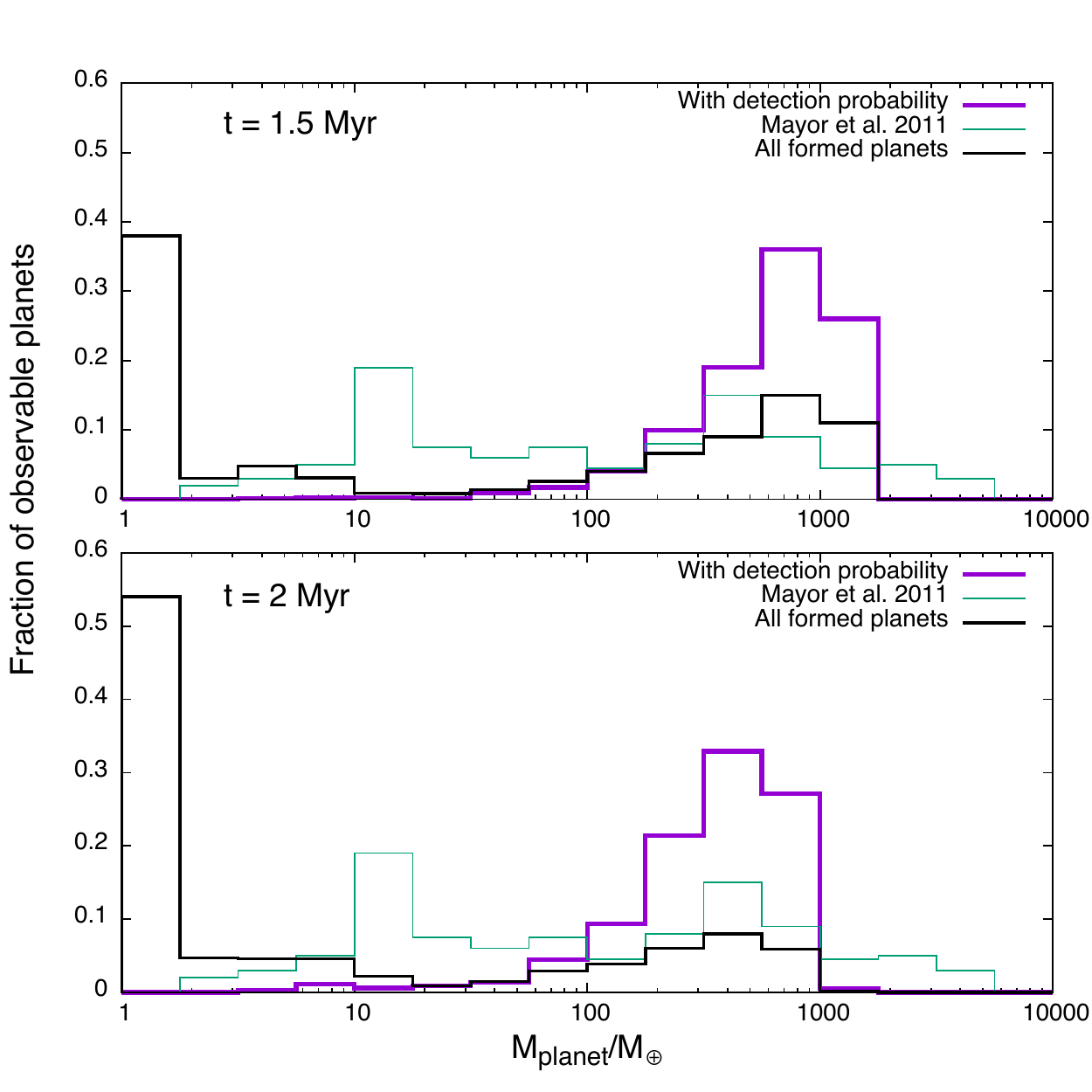}
  \caption{Mass function of our simulations using the B15b approach compared with the results presented in Mayor et al. 2011 (for planets with $a_{\rm final} > 0.1$ AU). The upper graph represents the simulations with embryos starting at $\rm t = 1.5 $ Myr of the evolution of the disc and the lower one the embryos starting after $\rm t = 2$ Myr. The green line represents the results presented in Mayor et al. 2011, the black line shows all the planets we form beyond 0.1 AU and the purple one indicates the planets considered as observed once the detection probability is applied.}
  \label{Massfunction_multiplot}
\end{figure}

In order to further compare our results with observations, we apply on our sample of planets the detection probability given by Mayor et al. 2011 in their Fig. 6.
To proceed, the masses of the formed planets are first represented as a function of their period (see Fig. \ref{multiplot_MvsP_1.5} (top graph) and \ref{multiplot_MvsP_2}). 
To be consistent with Mayor et al. 2011, we limit the period (in days) to $10'000$. 
We then apply their detection propability and determine which planet would be statistically detected and considered as observed (see Fig. \ref{multiplot_MvsP_1.5} bottom graph).  
We note that for $Z_{\rm tot} = 0.011$ (meaning $Z_{\rm peb} = 0.01$ and $Z_{\rm dust} = 0.001$ and represented by black dots) the planets are too small in mass to be detected.
Since we observe planets with radial velocity higher than 20 m/s to compare with Johnson et al. 2010, we do not detect the small planets interior to 1 AU as well.
The pile up we see in both figures on the left side represents the planets that might have migrated into the star and therefore, that are not observable. We compare the mass distribution from our models and Mayor et al. 2011 observations in Fig. \ref{Massfunction_multiplot}. 
We only take into account the data of Mayor et al. 2011 for planets with final location beyond 0.1 AU to be consistant with the inner edge of the disc in our model.

Observations (green line) show a high fraction of planets with masses around $10$ $\rm M_{\oplus}$ that our simulations (black line and purple line) do not reproduce.  
The black line takes all the planets beyond 0.1 AU into account. 
It shows a high fraction of Earth mass planets. 
However, essentially all of them are not observable (purple line) as their mass is too small given their location for them to be detected. 
Regardless of the starting time applied, only giant planets are detected.
However the masses of these planets differ between the two starting times: more massive planets form when embryos are placed after 1.5 Myr and less massive planets form when starting after 2 Myr. 
This can be explained by the fact that planets starting at 1.5 Myr have more time to grow.\\

\section{Internal structure for the planetary atmosphere}
\label{Differentinternalstructure}

In the B15b model the internal structure is not computed: once the planet has reached $\rm M_{\rm iso}$, a contraction phase begins as long as $M_{\rm env}$ is smaller as $M_{\rm core}$. Then a rapid gas accretion phase starts following Machida et al. 2010 (see section \ref{pebble_accretion}).
Therefore the accretion rate onto a small mass planet can be very large.
In reality, if $\rm M_{\rm iso}$ is small, once it is reached, the contraction of the envelope produces some luminosity that needs to be taken into account and that slows down gas accretion. 
We thus decided to compute the internal structure of the planet to assess its importance. Instead of using the values given in B15b, we now solve the following equations:

\begin{equation}
\frac{dm}{dr}= 4 \pi r^2 \rho,
\label{massconservation}
\end{equation}

\begin{equation}
\frac{dP}{dr} = - \frac{Gm}{r^2} \rho,
\label{evolutionofpressurewithdepth}
\end{equation}

and 
\begin{equation}
\frac{dT}{dr}= \frac{T}{P} \frac{dP}{dr} \nabla,
\label{energytransfer}
\end{equation}
which represent the mass conservation, the equation of hydro-equilibrium and energy transfer respectively (Alibert 2016). $\nabla = \frac{d \rm ln(T)}{d \rm ln(P)} = \rm min(\nabla_{\rm ad}, \nabla_{\rm rad})$ where
\begin{equation}
\nabla_{\rm ad} = \frac{\partial \rm ln(T)}{\partial \rm ln(P)} ,
\end{equation}
\begin{equation}
\nabla_{\rm rad}  = \frac{3}{64 \pi \sigma G} \frac{\kappa l P}{T^4 m},
\end{equation}
with $l$ being the luminosity of the planet computed by energy conservation and including the solid accretion luminosity, the gas contraction luminosity and the gas and core accretion luminosity (Mordasini et al. 2012, Alibert et al. 2013). $\kappa$ is the full interstellar opacity\footnote{A discussion on the opacity is provided in section \ref{effectofopacity}}.
Computing the internal structure at each timestep by solving the previous equations, the accretion rate onto the planet is automatically determined by the difference in the envelope mass at two consecutive timesteps (Alibert et al. 2005).\\

\begin{figure}
\hspace{0cm} \includegraphics[height=0.35\textheight,angle=0,width=0.35\textheight]{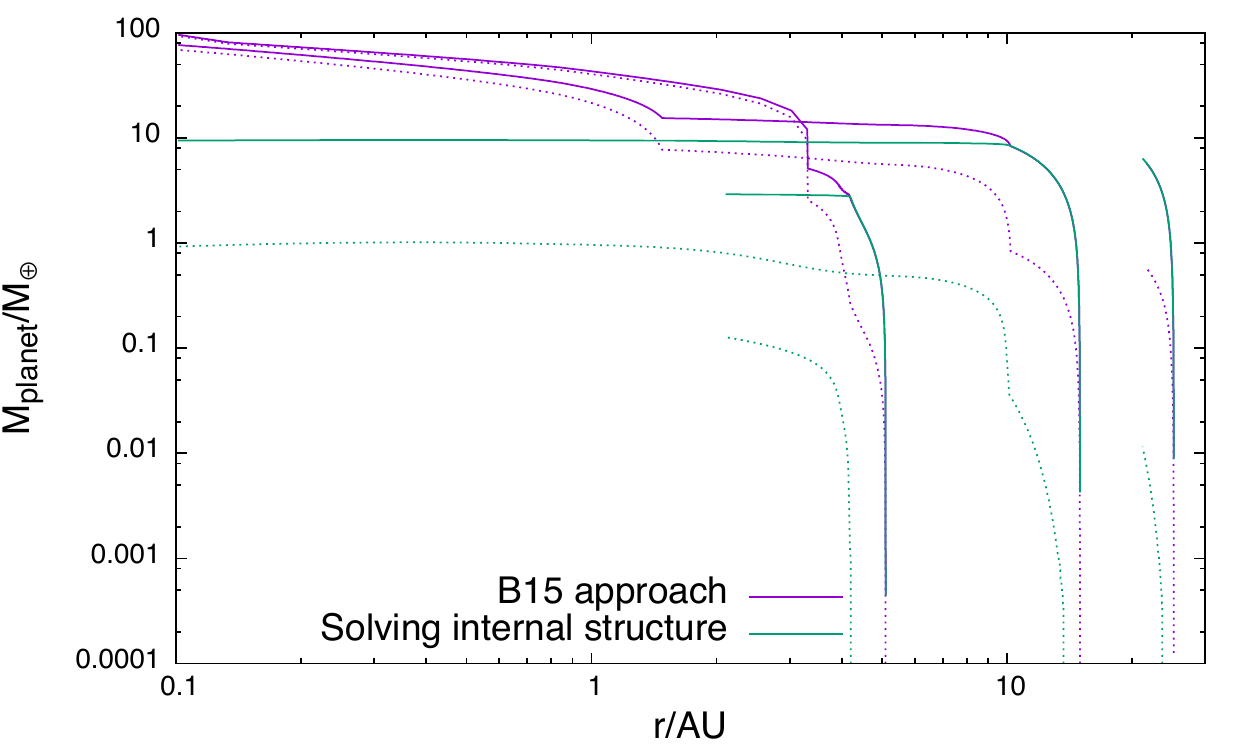}
  \caption{Growth tracks for planet starting at different locations. The purple curves show the growth of the planets using B15b approach and the green curves express the results solving the internal structure equations given in section \ref{Differentinternalstructure}. The dotted lines represent the envelope mass of each planet for both cases.}
  \label{Tracks_2internalstruc_comparison}
\end{figure}

The influence of the internal structure can be seen in Fig. \ref{Tracks_2internalstruc_comparison} which shows growth tracks for different starting locations. 
These tracks are similar for both approaches (B15b approach and solving the internal structure equations for the atmosphere) until the planets reach the isolation mass. 
Below $\rm M_{\rm iso}$ the accretion rate of solids is very high, therefore there is little gas. 
Thus how this little amount of gas is computed has no real influence when we look at the total planetary mass. 
Planets starting at large distances don't grow massive enough for internal structure to affect the formation tracks. 
Finally, it is for the planets starting between 5 and 15 AU that the internal structure effects are the most noticeable.
This comes from the fact that B15b assumes a rapid gas accretion once the isolation mass is reached, independently of the mass. 
Focusing on the planet starting at 5 AU, the isolation mass is $\sim 2.5$ $\rm M_{\oplus}$. 
Using B15b approach, once $M_{\rm iso}$ is reached, the accretion is huge and a final mass of $\sim 95$ $\rm M_{\oplus}$ is obtained, while computing the internal structure produces a planet with a mass of $\sim 2.5$ $\rm M_{\oplus}$.

\subsection{Comparison with observations}
\label{Differentinternalstructure_comparisonwithobs}

\begin{figure}
\hspace{0cm} \includegraphics[height=0.35\textheight,angle=0,width=0.35\textheight]{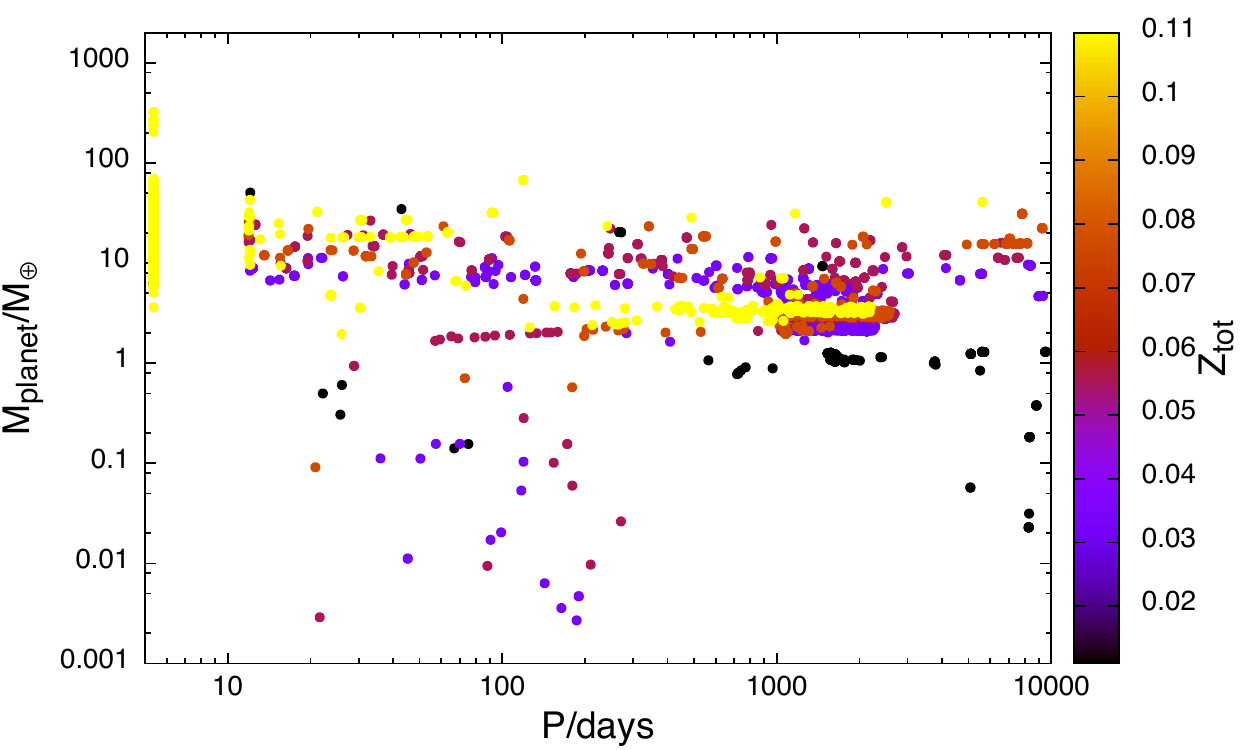}
  \caption{Mass of the planets as a function of their period for starting time of the embryos of $\rm t = 1.5$ Myr. These results include the internal structure as presented in section \ref{Differentinternalstructure}.}
  \label{multiplot_MvsP_1.5_otherint_struc}
\end{figure}

\begin{figure}
\hspace{0cm} \includegraphics[height=0.35\textheight,angle=0,width=0.35\textheight]{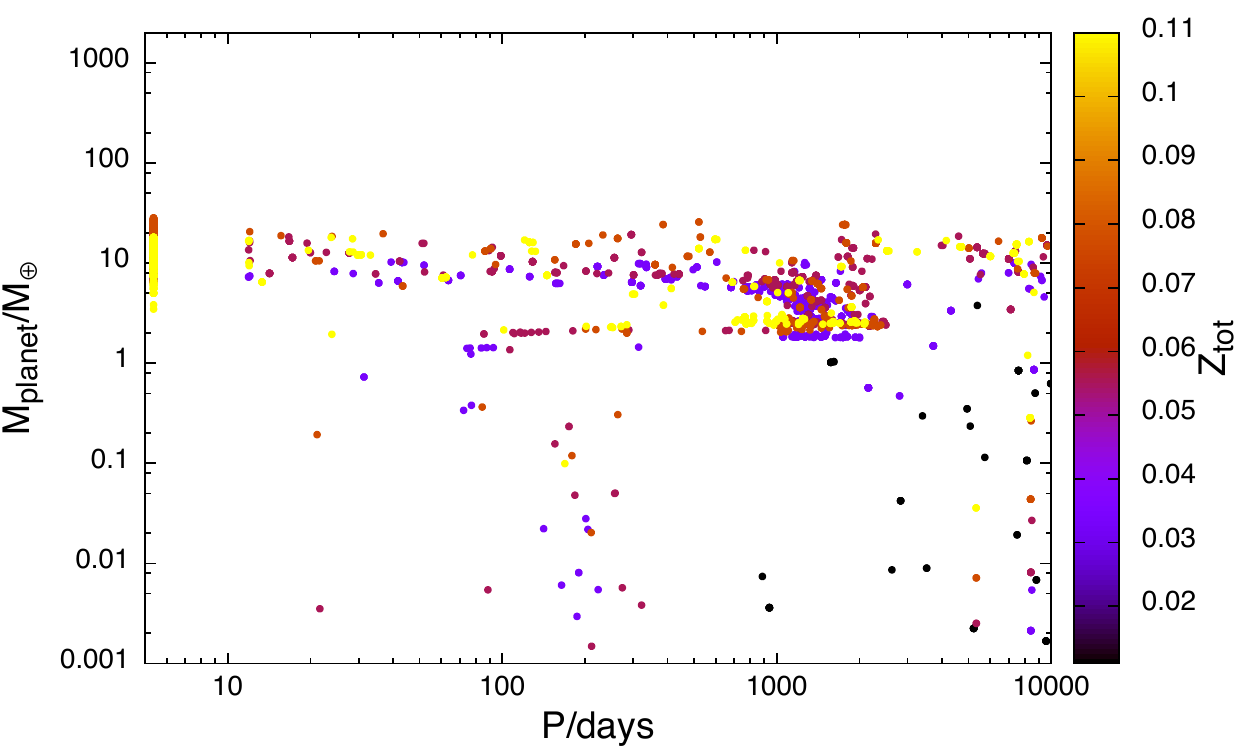}
 \caption{Same as Fig. \ref{multiplot_MvsP_1.5_otherint_struc} but for starting time of the embryos of $\rm t = 2$ Myr. The internal structure equations are computed here as well.}
  \label{multiplot_MvsP_2_otherint_struc}
\end{figure}

A whole new set of simulations is computed with the solving of the internal structure equations while keeping all other parameters constant. \\
The mass of the formed planets is given as a function of orbital period for starting times of the embryos of 1.5 and 2 Myr (see Fig. \ref{multiplot_MvsP_1.5_otherint_struc} and \ref{multiplot_MvsP_2_otherint_struc}) and it can again be observed that the higher the amount of solids, the bigger the planets, like we concluded with the B15b approach (Fig. \ref{multiplot_MvsP_1.5} and \ref{multiplot_MvsP_2}). 
Comparing Fig. \ref{multiplot_MvsP_1.5_otherint_struc} and \ref{multiplot_MvsP_2_otherint_struc} with the two former plots, we see that the masses of the planets are much smaller.
This is due to the fact that solving the equations for the internal structure results is an additional energy source related to envelope contraction which will slow down gas accretion.
Therefore small cores that have reached the isolation mass cannot accrete a huge amount of gas efficiently.\\
The internal structure also decreases the amount of planets remaining beyond 0.1 AU from the star. 
The formed planets are not big enough to open a gap in the disc and to migrate in type II. 
They thus migrate with type I migration which is rapid and makes most of them fall into the star. \\

\begin{figure}
\hspace{0cm} \includegraphics[height=0.35\textheight,angle=0,width=0.35\textheight]{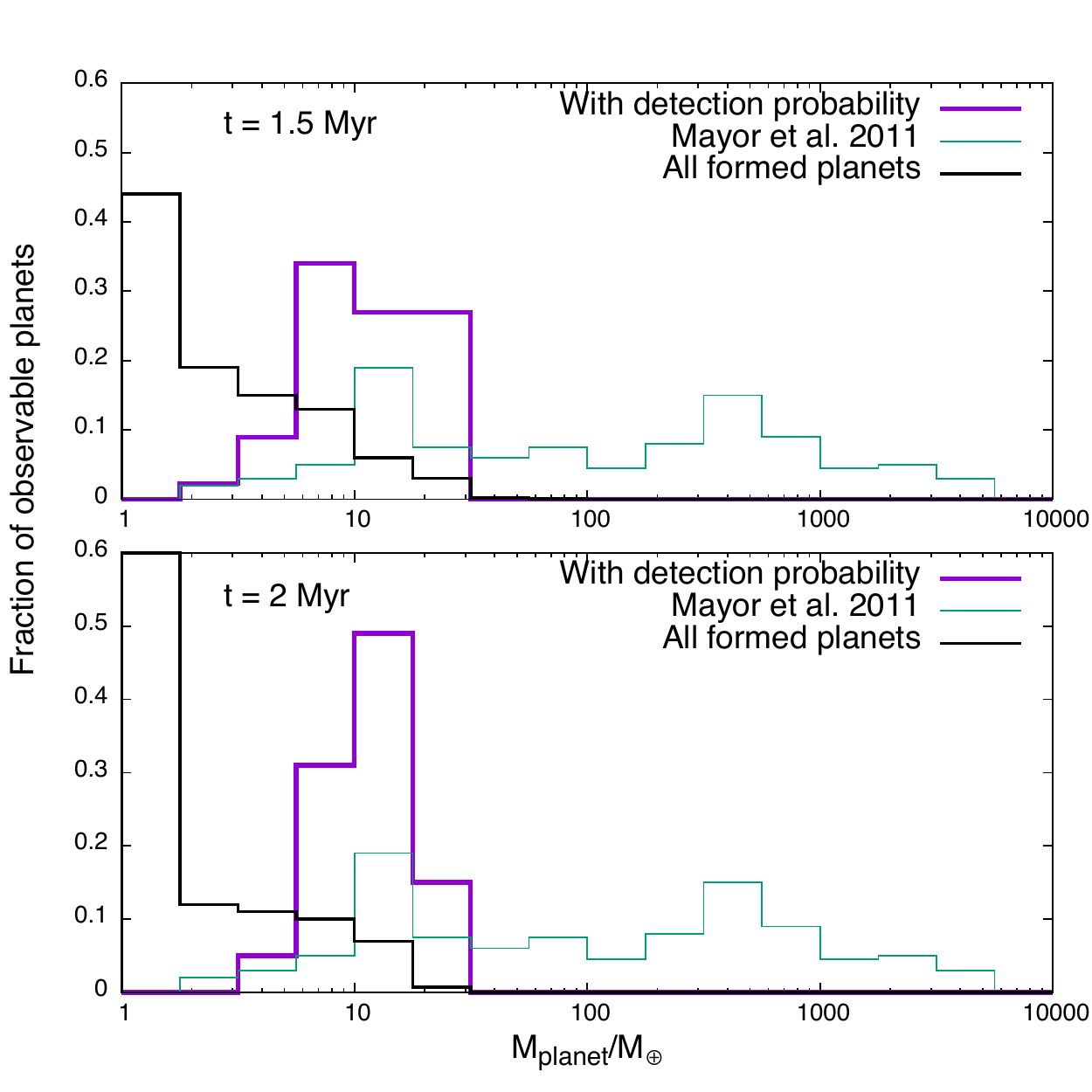}
  \caption{Mass function of the planets having an internal structure compared with the results presented in Mayor et al. 2011 (for planets with $a_{\rm final} > 0.1$ AU). The upper graph represents the simulations with embryos starting at $\rm t = 1.5$ Myr  of the evolution of the disc and the lower one the embryos starting after $\rm t = 2$ Myr.}
  \label{Massfunction_multiplot_otherinternalstructure}
\end{figure}

Focusing on the mass of the formed planets, the difference with the B15b approach is not negligible.
This is highlighted in Fig. \ref{Massfunction_multiplot_otherinternalstructure} which shows the fraction of observable planets as a fonction of their mass, for final semi-major axis beyond 0.1 AU.
A high amount of Neptune mass planets and a total absence of giants are seen here. 
This result is completely opposite to the one obtained using B15b approach (see Fig. \ref{Massfunction_multiplot}).


\subsection{Effect of the starting time of the embryo}

As we said in section \ref{pebble_accretion}, we assume that the embryos in the disc have already formed through streaming instability, at a given time that is a free parameter of the model. 
This initial time has an impact on the type of formed planets but also on the amount of planets remaining beyond 0.1 AU (see Table \ref{Tableau_pourcent}).
Even though the percentage of surviving planets is quite low when assuming early formation times for the embryos (0, 0.5 Myr), we decide to compare the type of formed planets for different starting times computing the internal structure and to compare those results with the ones for the B15b approach.\\

\begin{figure}
\hspace{0cm} \includegraphics[height=0.35\textheight,angle=0,width=0.35\textheight]{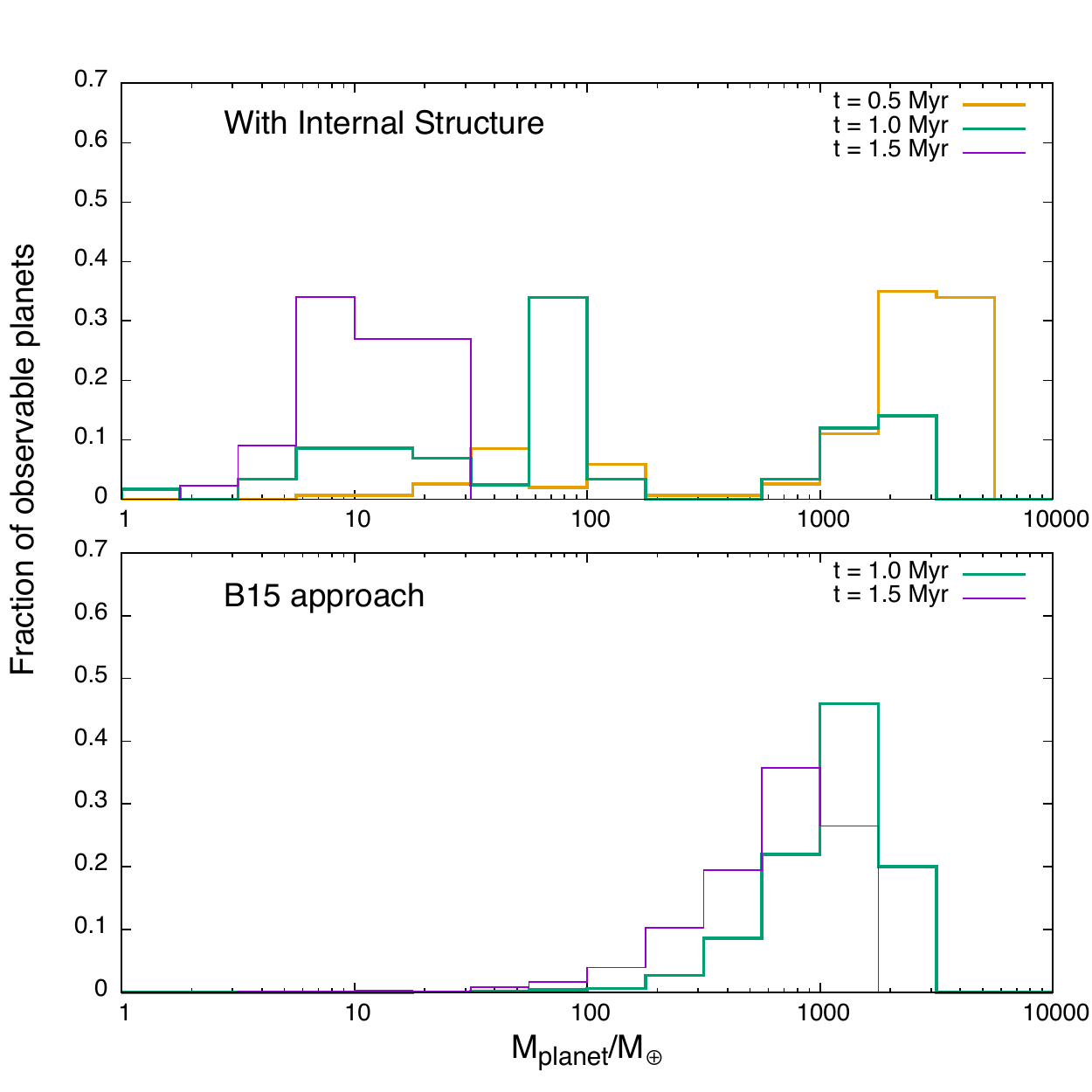}
  \caption{Comparison of mass functions for two different models and starting times (the detection probability of Mayor et al. 2011 is applied for both graphs). The upper plot represents the model computing the internal structure equations. It shows that decreasing the starting time of the embryos, for instance from 1.5 Myr to 1 Myr, allows the appearance of some giants. The bottom plot represents the B15b approach. We see here that decreasing the starting time of the embryos does not have a huge impact since mainly giant planets formed with this model.}
  \label{multiplot_massfunction_allstartingtimes}
\end{figure}

Fig. \ref{multiplot_massfunction_allstartingtimes} (upper plot) shows the mass distribution of planets formed as a function of the embryo starting time.
We see that an early start favours the formation of more massive planets.
We concluded in section \ref{Differentinternalstructure_comparisonwithobs} that solving the internal structure equations mainly formed Neptune mass planets and practically no giants. 
With the upper plot of Fig. \ref{multiplot_massfunction_allstartingtimes} we see that inserting our embryos earlier in the disc evolution, for instance at $\rm t_{\rm ini} = 1$ Myr, would give them more time to grow and thus allow to form giant planets. \\
For comparison, in the B15b approach the mass distribution of the planets remains quite similar for $\rm t_{\rm ini} = 1$ Myr or $1.5$ Myr (see Fig. \ref{multiplot_massfunction_allstartingtimes} (bottom plot)). \\

The results for $\rm t_{\rm ini} = 0.5$ Myr computing the internal structure need to be considered with caution due to the high amount of planets ending into the star (see Table \ref{Tableau_pourcent_WITH_IS}). 
The percentages shown in Table \ref{Tableau_pourcent_WITH_IS} can be compared with Table \ref{Tableau_pourcent} which labelled the amount of planets falling into the star for the B15b approach.
We see that in both cases, the earlier the embryos are inserted, the more planets fall into the star.\\

\begin{table}
\centering
\caption{\label{Tableau_pourcent_WITH_IS} Percentages of planets lost in the star (closer than 0.1 AU) according to the different starting times of the seed for the model with an internal structure.}
\begin{tabular}{|c|c|} 
\hline
Starting times & \% of planets ending in the star \\
\hline
0.5 Myr & 78 \% \\
\hline
1 Myr & 67 \% \\
\hline
1.5 Myr & 53 \% \\
\hline
2 Myr & 32 \% \\
\hline
\end{tabular}
\end{table}

\subsection{Effect of the opacity}
\label{effectofopacity}

Another alternative to form giants is by reducing the opacity of the planet atmosphere. 
This opacity can be separated into a gas and a grain component.
The gas part remains unchanged.
In our nominal case the grain part is provided by the full interstellar grain opacity (Lin \& Bell, 1994).
We now apply a reduction factor $\rm f_{\rm opa}$ on the grain opacity which will influence the envelope mass and the planet formation timescale (Mordasini et al. 2014).
The reason behind this reduction is that, once $M_{\rm iso}$ is reached, the velocity of the gas outside the planet's orbit becomes super-Keplerian, halting the drift of pebbles. 
Therefore the gas accreted by the planet will contain a reduced amount of solids (Bitsch \& Johansen 2016).
The resulting consequence is that, if $\rm f_{\rm opa}$ is used, the grain opacity is then smaller leading to a shorter formation timescale (Ikoma et al. 2000) and a bigger envelope mass (Mordasini et al. 2014).

\begin{figure}
\hspace{0cm} \includegraphics[height=0.35\textheight,angle=0,width=0.35\textheight]{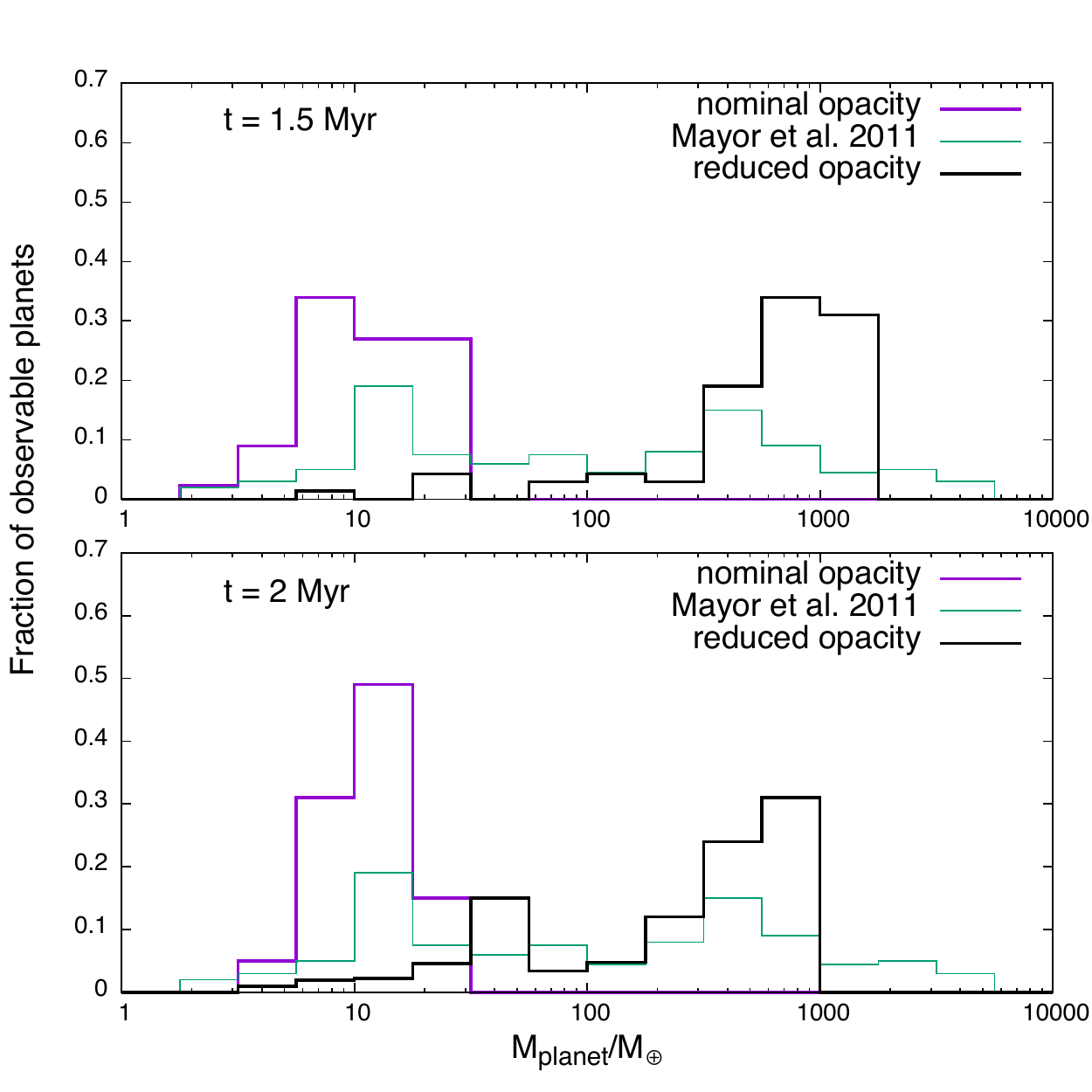}
  \caption{We compare here the type of formed planets using the nominal case grain opacity (which is the full interstellar one) and a reduced one (by a factor $\rm f_{\rm opa} = 0.01$) for the model computing the internal structure. The upper plot is for a starting of $\rm t = 1.5$ Myr and the bottom one for $\rm t = 2 $ Myr. The green line represents the results of Mayor et al. 2011 for planets beyond 0.1 AU for comparison. }
  \label{massfunction_smalleropacity}
\end{figure}

Fig. \ref{massfunction_smalleropacity} confirms this tendency.
It shows that using the model computing the internal structure and reducing the opacity will lead to a change in the mass distribution of planets.
The fraction of planets with masses larger than 20 $M_{\oplus}$ is significant and a large amount of giant planets are observable for both starting times. 
This is in better agreement with observations (Mayor et al. 2011) and we therefore use this model for the following section (section \ref{metallicity}).\\

\begin{figure}
\hspace{0cm} \includegraphics[height=0.35\textheight,angle=0,width=0.35\textheight]{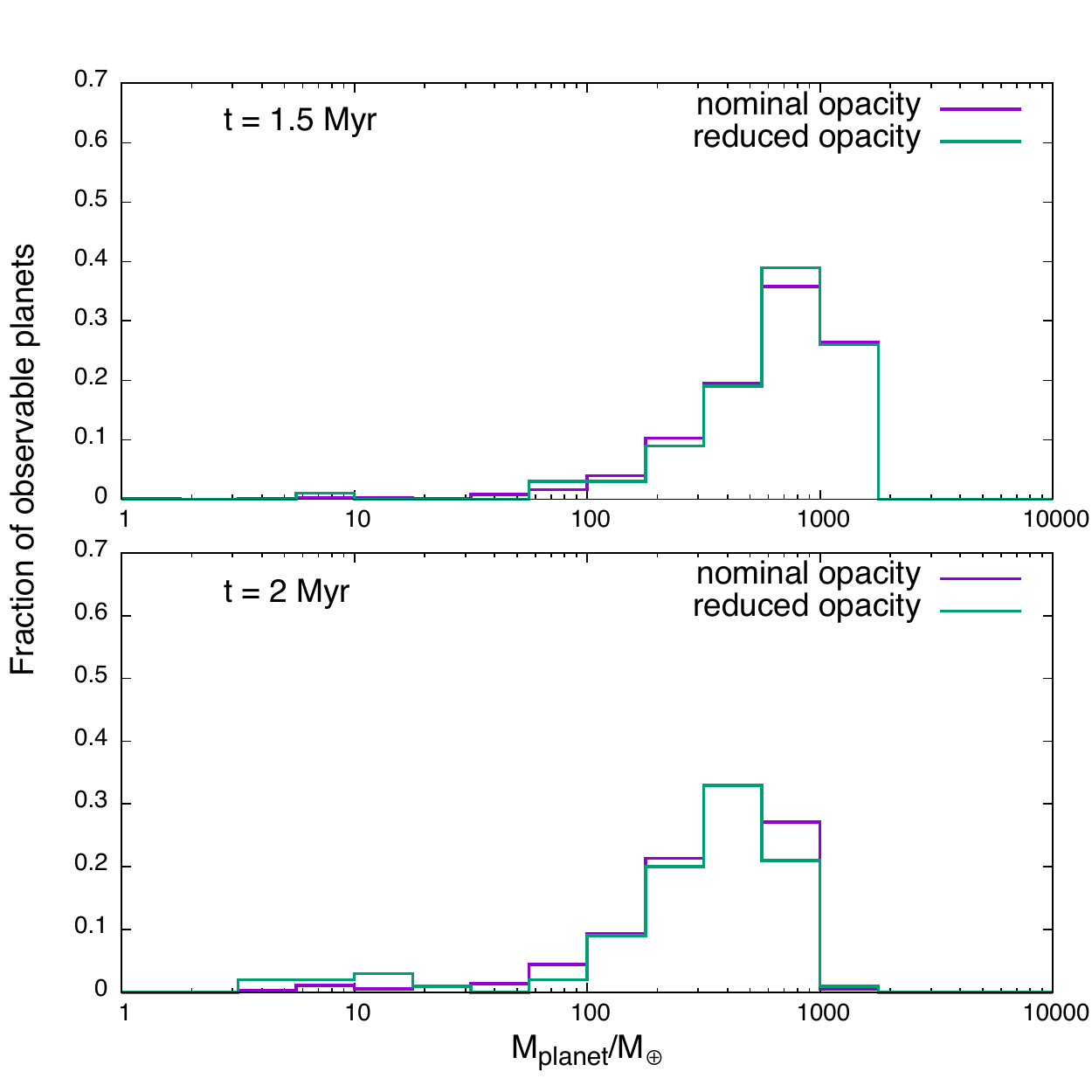}
  \caption{We compare here the B15b approach for the regular fixed opacity ($\kappa = 0.05$) given in their model and a reduced one by a factor $\rm f_{\rm opa} = 0.01$ leading to $\kappa = 0.0005$. The upper plot is for a starting of $\rm t = 1.5$ Myr and the bottom one for $\rm t = 2 $ Myr. }
  \label{massfunction_B15_smalleropacity}
\end{figure}

Reducing the opacity for the B15b approach we see (Fig. \ref{massfunction_B15_smalleropacity}) that the mass distribution of formed planets does not differ.
This leads us to the conclusion that the main difference in the types of formed planets between the two models (B15b approach and computing the internal structure) comes from the internal structure and not the opacity.

\section{Metallicity effect}
\label{metallicity}

\begin{figure}
\hspace{0cm} \includegraphics[height=0.35\textheight,angle=0,width=0.35\textheight]{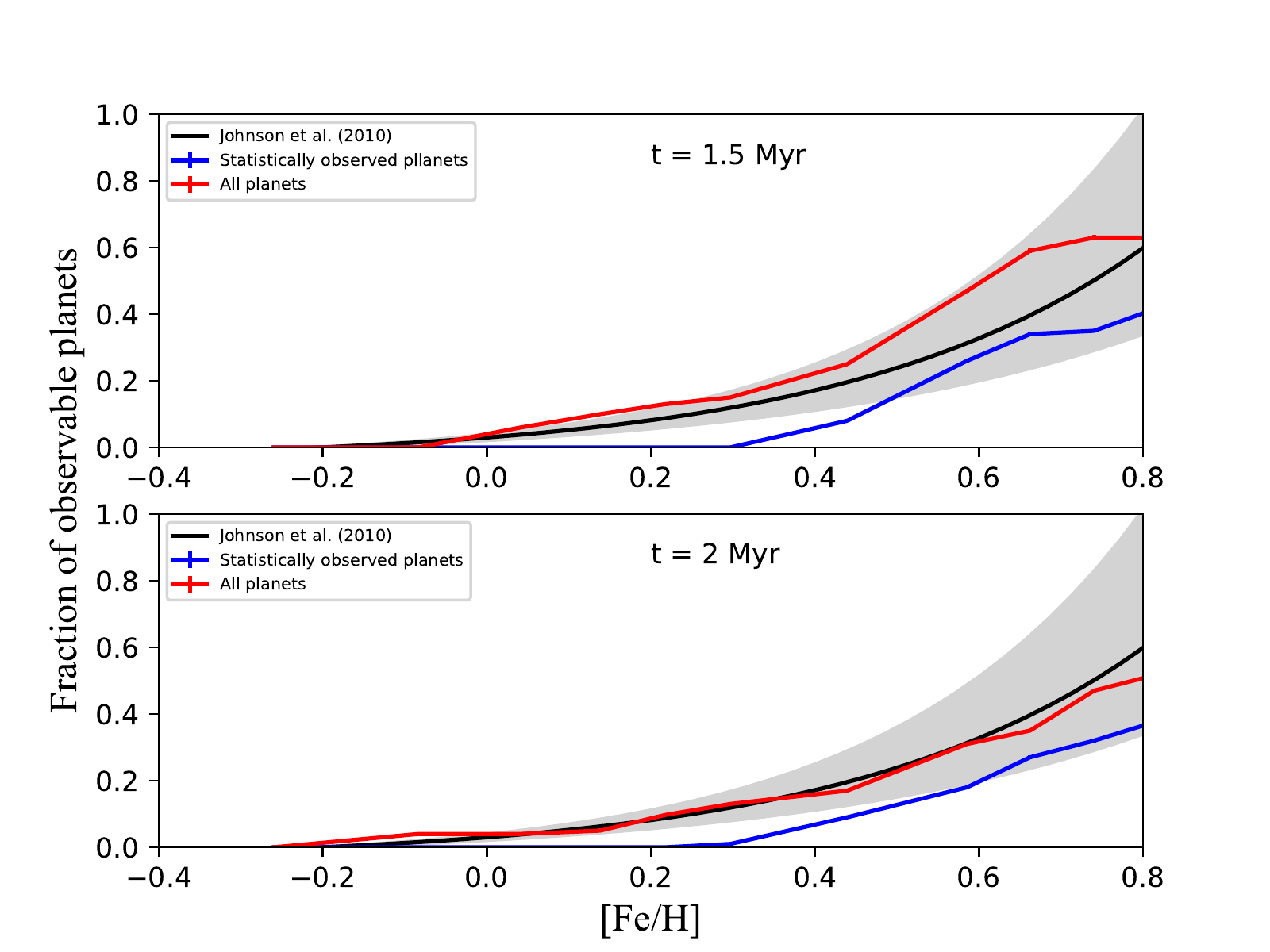}
  \caption{Fraction of planets with radial velocity higher than 20 m/s as a function of the metallicity for the model computing the internal structure with a reduced opacity. The top plot is for seeds starting after 1.5 Myr of the disc evolution and the bottom one for $\rm t_{\rm ini} = 2$ Myr. The red line shows the percentages of planets with a radial velocity higher than 20 m/s. The blue line is the percentages of planets that has a radial velocity higher than 20 m/s and that can be observable according to the detection probability of Mayor et al. 2011. The black line represents the observation fit given by Johnson et al. 2010 (Eq. \ref{Johnson}) and the grey area the uncertainty on this fit. }
  \label{KrvIS}
\end{figure}

We now want to discuss the well established metallicity effect for our internal structure model with a reduced opacity. 
We chose  the solar value $ Z_{\rm tot} = 0.02$ to be the reference value for which $\rm[Fe/H] = 0$ (Anders \& Grevesse, 1989). 
We vary  $Z_{\rm tot} = Z_{\rm peb} + Z_{\rm dust}$ and run simulations for each starting time of the seeds. 
In Fig. \ref{KrvIS} we plot the fraction of planets that have a radial velocity higher than 20 m/s as a function of the metallicity of the star. 
This radial velocity corresponds to the value used by Johnson et al. 2010, with which we compare our results. 
As given in Johnson et al. 2010 the fraction of stars with giant planets as a function of metallicity, based on their radial velocity measurements, follows:
\begin{equation}
f = (0.07 \pm 0.01) \cdot (M_{\rm star} / M_{\odot})^{(1.0 \pm 0.3)} \cdot 10^{(1.2 \pm 0.2) \rm[Fe/H]},
\label{Johnson}
\end{equation}
where, in our case, $M_{\rm star} = M_{\odot}$.
 In our figure, the red line takes all formed planets  into account. 
 The percentages are then higher than the blue curve, which represents the planets that are statistically observed, once the detection probability of Mayor et al. 2011 is applied. 
 Eq. \ref{Johnson} (Johnson et al. 2010) is given by the black curve and matches well with our simulations.\\

The general tendency that can be observed in Fig. \ref{KrvIS} is that the amount of giant planets increases with metallicity. For metallicities up to $\rm[Fe/H] = 0.3$, no planet with radial velocity higher than 20 m/s remains beyond 0.1 AU, and thus can be observable. 
This is related, as mentionned in section \ref{Formation_tracks}, to the inverse correlation between dust amount and migration rate.
Indeed, focusing on a smaller metallicity (for exemple $\rm [Fe/H] = 0.2$) and comparing the red and blue curves, we see that a fraction of formed planets (red curve) may be lost in the star (difference between red and blue).\\ 

\begin{figure}
\hspace{0cm} \includegraphics[height=0.35\textheight,angle=0,width=0.35\textheight]{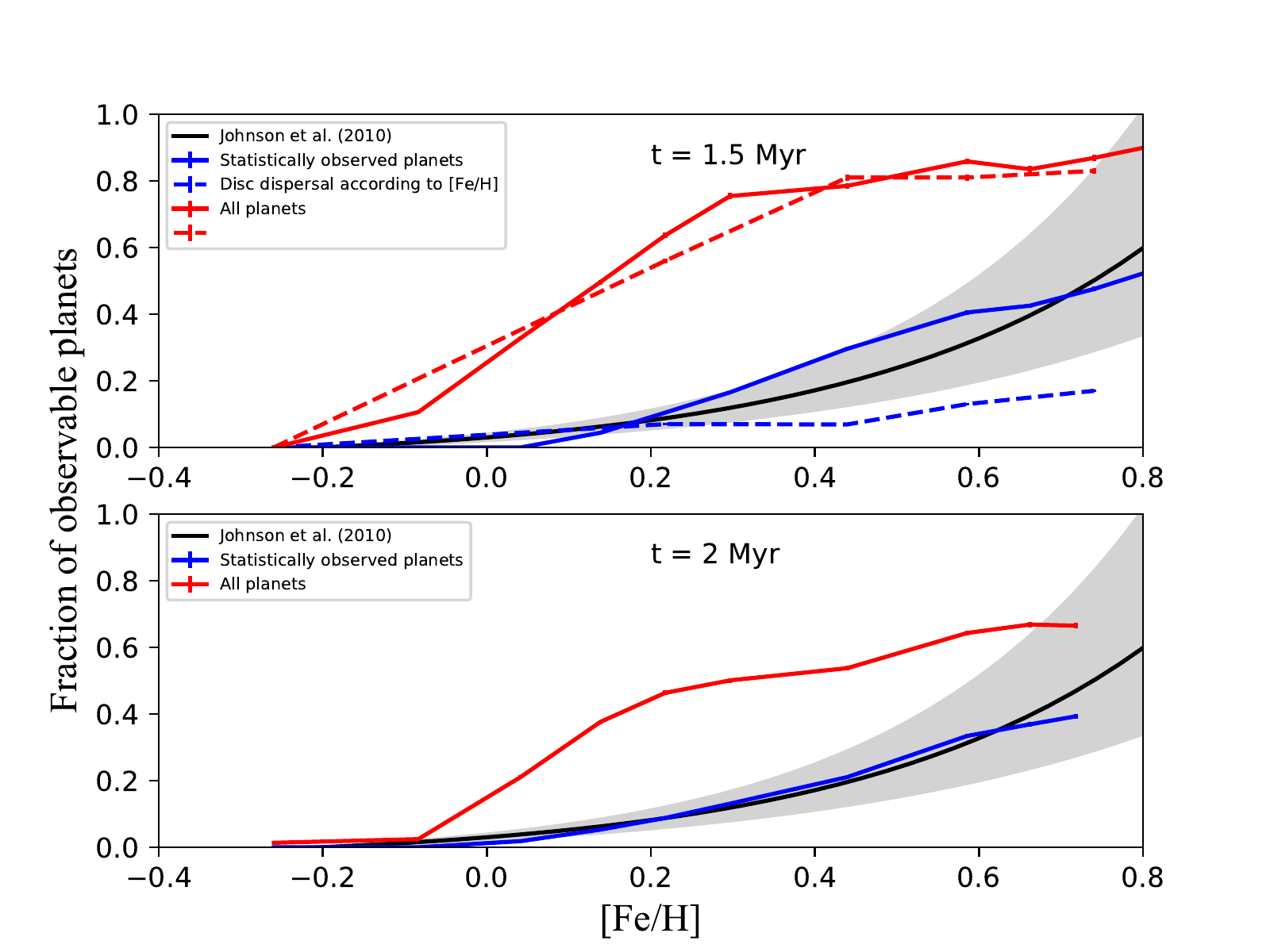}
  \caption{Same as in Fig. \ref{KrvIS} but with the B15b approach. The dotted lines take the disc dispersal according to [Fe/H] into account.}
  \label{KrvB15}
\end{figure}

Fig. \ref{KrvB15} is similar to Fig. \ref{KrvIS} but using the B15b approach. 
The general tendency already observed in Fig. \ref{KrvIS} that the higher the metallicity, the more massive the planets, is also verified with this model (see also Ndugu et al. 2017). 
The main difference comes from the fact that, when all the planets are taken into account, a huge fraction of them have a radial velocity higher than 20 m/s. 
Thus the blue curves (for observable planets) fit the curve given by Johnson et al. 2010 but the red one (all planets) shows much higher percentages. This indicates that, using the B15b approach, a lot of giants form (since gas accretion is huge once $\rm M_{\rm iso}$ is reached) but most of them are lost into the star. \\

The dotted line in the upper plot of Fig. \ref{KrvB15} includes the effect of the metallicity on the disc lifetime. 
Our model has a disc lifetime that is randomly chosen between 2 and 5 Myr (Haish et al. 2001b, see section \ref{Initialconditions}) and does not take the impact of the metallicity into account.
Based on Yasui et al. 2010, a low disc metallicity may cause a rapid disc dispersal via photoevaporation.
In order to test the impact of the metallicity on the disc lifetime we set the solar metallicity disc lifetime to 3 Myr and multiply by $10^{[Fe/H]}$ in order to get the other lifetimes.
For high metallicites (from $\rm Z_{\rm tot}$ = 0.033 to 0.11) the disc lifetimes go from 4.95 Myr up to 16.5 Myr, leading to larger planets but more than 90\% of them migrate to the inner egde of the disc. 
The surviving planets are contained within 0.2 AU.
For low metallicity ($\rm Z_{\rm tot}$ = 0.011), the disc lifetime is 1.65 Myr and planets don't grow much since they are inserted in the disc after 1.5 Myr. 
They therefore remain with masses of the order of $10^{-3}$ $\rm M_{\oplus}$, near their initial location.
Despite this difference in the population, the amount of planets with a radial velocity higher than 20 m/s as a function of the metallicity looks similar to what we obtain with the nominal model (see Fig. \ref{KrvB15}, upper plot). The higher the metallicity, the higher the amount of planets with radial velocity larger than 20 m/s. The percentages for planets beyond 0.1 AU (dotted blue line) seems however less influenced by the increase in metallicity, but only a very few of the planets remain beyond the inner edge of the disc.

We can conclude that both models show an impact of the metallicity.

 \section{Discussion and conclusion}
 \label{conclusion}
 
In this work we provide an updated pebble-based planet formation model. 
We use the disc model given by Bitsch et al. 2015a and first test our computations by intending to reproduce their results. 
Aiming at this we noticed that the equation used to compute the flux of pebbles in B15b is not correct, as confirmed later on by B. Bitsch (private communication, Bitsch et al. 2017). 
Using the  correct equation translates in the formation of very small mass planets, at least when the fraction of the total amount of solids taking part in the formation of pebbles is the same as in B15b ($Z_{\rm peb} = 2/3 \cdot Z_{\rm tot}$). 
Adopting the correct pebble flux equation therefore requires assuming that a larger fraction of the heavy elements takes part in the formation of pebbles ($Z_{\rm peb} / Z_{\rm tot} = 0.9$) in order to form massive enough planets.\\

Using this approach we study the impact of the amount of pebbles ($Z_{\rm peb}$) as well as the amount of dust ($Z_{\rm dust}$).
The latter impacts not only the disc structure but also the growth of the planets through the isolation mass. 
$Z_{\rm peb}$ influences the surface density of pebbles and thus the planet growth. 
We show that increasing the density of pebbles favours the formation of massive planets. 
However the amount of dust should not be too small otherwise migration is too efficient and planets fall into the star.
We therefore see that the changes in the disc properties (through $Z_{\rm tot}$ and thus $Z_{\rm dust}$) lead to a large diversity of outcomes. This is similar to Ida et al. 2016 where the boundaries in the disc (affected by the disc properties) have a strong influence on the results. \\

Using a population synthesis approach we then compute simulations and compare our results with observations (Mayor et al. 2011).
The latter indicates a high fraction of Neptune mass planets that we do not reproduce. 
Indeed our mass distribution is such that most of the detectable planets are giant planets.
This significant amount of giants come from the fact that in the B15b model, there is no internal structure.
Indeed once the isolation mass is reached a huge amount of gas is accreted.
In reality, once the planet reaches $\rm M_{\rm iso}$, the envelope contracts, producing some luminosity that will slow down gas accretion.\\

Computing the internal structure for the planetary atmosphere we produce another full set of simulations. 
We obtain a high amount of Neptune mass planets and a lack of giant planets.
To recover an acceptable number of giant planets, we consider two different parameters: the starting time of the seeds and the opacity of the planetary atmosphere.\\
The starting time may indeed have a considerable impact on the outcome of the simulation, as shown by Chamber 2016.
Studying the outcomes of planetary systems, they show that the embryos should have a mass between $10^{-3}$ to $10^{-2}$ $M_{\oplus}$ at 1.5 Myr and 2 Myr of their evolution to accrete efficiently and eventually grow to giant planets.  
In our model, we do not compute the growth of the embryos but insert them at later times (1.5 Myr and 2 Myr) with masses that are consistant with their results.
Therefore the masses the embryos have at a given time influence the final planetary architecture for multiple-planets systems (Chambers 2016, Levison et al. 2015) as well as for single planet cases (this work).
We highlight that inserting our embryos earlier in the disc evolution, for instance at $\rm t_{\rm ini} = 1$ Myr, would give them more time to grow and help forming giant planets.
Reducing the opacity also increases the amount of giants and the resulting type of formed planets is in good agreement with observations (Fig. \ref{massfunction_smalleropacity}).\\
We thus use this last model to discuss the metallicity effect. 
We show that the amount of giant planets increases with the metallicity and that the B15b approach also reproduces this effect.\\

The main difference between the B15b approach and our model (with the internal structure and a reduced opacity) is the fact that we compute an internal structure for the planetary atmosphere. 
We conclude that computing the internal structure and reducing the opacity is more realistic and reproduces better the mass distribution of the observed planets (Mayor et al. 2011).
However the replenishment of the envelope with the surrounding gas disc is not taken into account here which may lead to a more sophisticated model (Lambrechts \& Lega 2017, Brouwers et al. 2017).\\
Finally we should also keep in mind that our simulations are performed with one planet per disc and thus the eventual effects of several planets growing in the same disc are not taken into account, which we leave for future work (Coleman et al. in prep.), as well as long-term evolution, which is essential in order to compute the present day radius of planets. In addition, for planets close enough to the star, and of intermediate mass, evaporation needs to be taken into account (Mordasini et al. 2012).

\acknowledgements

This work has been carried out within the frame of the National Centre for Competence in Research PlanetS supported by the Swiss National Science Foundation. The authors acknowledge the financial support of the SNSF. We thank Gavin Coleman for helpful comments as well as Yuhito Shibaike. The authors also would like to thank B. Bitsch and the referee for their comments.

\end{document}